\documentclass[reprint,aps,twocolumn,superscriptaddress]{revtex4-1}

\usepackage{amssymb,amsmath,epsfig,color, graphicx}   
\usepackage{subfigure, float, dcolumn, bm, epstopdf}
\usepackage{blindtext, natbib}
\usepackage{hyperref}
\usepackage{cleveref}

\let\oldcite=\cite 
\renewcommand{\cite}[1]{\textcolor[rgb]{0,0,1}{\oldcite{#1}}}

\let\oldref=\ref 
\renewcommand{\ref}[1]{\textcolor[rgb]{0,0,1}{\oldref{#1}}}

\begin{document}

\title{ Molecular dynamics approach for predicting release temperatures of noble gases in pre-solar nanodiamonds  }

\author{Alireza Aghajamali }
\email{alireza.aghajamali@curtin.edu.au }
\affiliation{Department of Physics and Astronomy, Curtin University, Perth, Western Australia 6102, Australia}
\author{Andrey A. Shiryaev}
\affiliation{Frumkin Institute of Physical Chemistry and Electrochemistry RAS, Leninsky pr.31 korp. 4, Moscow 119071, Russia}
\author{Nigel A. Marks }
\affiliation{Department of Physics and Astronomy, Curtin University, Perth, Western Australia 6102, Australia}


\begin{abstract}
Pre-solar meteoritic nanodiamond grains carry an array of isotopically distinct noble gas components and provide information on the history of nucleosynthesis, galactic mixing and the formation of the Solar system. In this paper, we develop a molecular dynamics approach to predict thermal release pattern of implanted noble gases (He and Xe) in nanodiamonds. We provide atomistic details of the unimodal temperature release distribution for He and a bimodal behavior for Xe. Intriguingly, our model shows that the thermal release process of noble gases is highly sensitive to the impact and annealing parameters as well as to position of the implanted ion in crystal lattice and morphology of the nanograin. In addition, the model elegantly explains the unimodal and bimodal patterns of noble gas release via the interstitial and substutional types of defects formed. In summary, our simulations confirm that
low-energy ion-implantation is a viable way for the incorporation of noble gases into nanodiamonds and we provide explanation of experimentally observed peculiarities of gas release.

\end{abstract}


\date{\today}

\maketitle

\section{Introduction}

Concentration of noble gases (NG) in primitive chondrites may greatly exceed terrestrial abundancies. Several NG components, different in chemical and isotopic composition, can be distinguished using step combustion experiments (see \cite{Ott-Geochemistry-2014} for review). A significant fraction of the anomalous components, i.e., those with isotopic composition markedly different from the solar, reside in nanodiamond (ND) grains, which are relatively abundant in some primitive chondrites with content as high as 1500~ppm. 

Step combustion and pyrolysis experiments allowed to distinguish three NG components present in NDs -- P3, HL and P6. Each of these components contain all five NGs (He, Ne, Ar, Kr, Xe); understanding of Xe is, arguably, less ambiguous than the other elements. The P3 and, possibly, the P6 components, are not dramatically different from the solar elemental and isotopic patterns, but still their composition cannot be derived from the solar by fractionation. 

The P6 component is volumetrically the least important and is
yet insufficiently characterized. At least a fraction of the P3 and P6-related
$^{129}$Xe is derived from the decay of $^{129}$I, implying the initial
presence of currently extinct iodine in NDs (e.g., Gilmour, \emph{et al.} \cite{Gilmour-GCA-2005}).
The HL component is characterized by marked excess in the relative abundance of
heavy (H) and light (L) isotopes. The heavy isotopes -- $^{134}$Xe and
$^{136}$Xe -- are produced by $r$-process only, or, at least, overwhelmingly.
The light isotopes -- $^{124}$Xe and $^{126}$Xe -- are formed via the
$p$-process. Historically, the only feasible explanation for the origin of the
HL component was a mixture of NGs produced in $p$- and $r$-processes in
supernovae explosions of type I \cite{Jorgensen-Nature-1988} or II
\cite{heymann1979xenon}. However, none of the current models of the
$r$-process fit the experimentally observed isotopic pattern and models with
variable degree of robustness were proposed \cite{ott_2013,
Ott-Geochemistry-2014}. In any case, isotopic patterns suggest that
NDs carrying the HL, and, likely also the P3 and P6 components, should
be of presolar origin. This view is also supported by the association of
isotopically anomalous Pt, Ba, Te and some other elements with ND-rich
fractions. Importantly, all attempts to separate individual carriers of the L and H
components failed, thus indicating thorough mixing of several independent
populations of NDs in presolar nebula or elsewhere. Since only approximately
$1$ out of $10^6$ ND grains contains a Xe ion, and all isotopic analyses
require billions of grains, the mixing is not that surprising. Existence of
several populations of NDs, possibly of different origin and/or stellar
sources, is also suggested by C and N isotopic composition
\cite{russell1996carbon}.

Step pyrolysis/oxidation experiments revealed complex patterns of NG release from NDs. For heavy NGs -- Ar, Kr, Xe, the gas release pattern is bimodal. Although some variations between meteorites with different thermal histories occur, in general the P3 component is released mostly at low temperatures (200--900) with a peak at 450--550$^\circ$C; the HL component peaks between 1300--1450$^\circ$C (overall range 1100--1600) and the P6 component evolve at yet higher temperatures. For Ne the bimodality is much less pronounced and for He it is probably absent, although its composition changes with the release temperature. 

The mechanism of incorporation of NGs into ND grains is a
subject of ongoing debate. In principle, three mechanisms may be responsible
for the introduction of an impurity atom into an ND grain: (1) growth process;
(2) diffusion and/or adsorption; (3) ion implantation. The largest atom ever
introduced into the diamond lattice purely by growth processes relevant for the
formation process of NDs in space -- plasma-assisted Chemical Vapour Deposition
-- is tin \cite{westerhausen2020controlled}. However, in contrast to NGs, Sn,
Ge, and other heavy elements belong to group IV of the periodic table as
well as C. Subsequently, the introduction of these elements is
crystallochemically feasible even despite large differences in atomic radii
with C. Diffusion in diamond is extremely sluggish \cite{koga2003diffusive},
which effectively rules out any diffusion-related processes at low
temperatures, relevant for meteoritic NDs. The surface adsorption of some
elements is possible, and several works suggest that the P3 component might be
trapped in near-surface, perhaps partly graphitised layers of the grains (e.g.,
Fisenko, \emph{et al.} \cite{fisenko2010nature}). However, ion implantation, where NGs are driven
into pre-existing NDs \cite{Lewis-Astro-1981, Lewis-Nature-1987,
Verchovsky-Science-1998, Gilmour-GCA-2005}, seems to be the only feasible
option \cite{Huss-Met-1994-I}. 

The implantation hypothesis was strongly supported by experimental work by Koscheev, \emph{et al.} \cite{Koscheev-Nature-2001} who implanted a low energy (700~eV) NG ions into synthetic NDs with diameters 4--5~nm. Pyrolysis measurements observed a single broad temperature distribution for He and Ne, and a bimodal distribution for Ar, Kr and Xe, corresponding closely to those found for meteoritic NDs \cite{Huss-Met-1994-I, Huss-Met-1994-II}. Similar studies by Verchovsky, \emph{et al.} \cite{Verchovsky-Abs-2000} also found a bimodal distribution, albeit in the latter case the high dose implantation most likely amorphised the ND grains, greatly complicating interpretation. However, physical mechanism of the bimodal gas release pattern remains unexplained and raised important questions whether the observed multimodality is real or results from experimental artifacts. For example, a single layer of ND grains was used as a target in Koscheev, \emph{et al.} \cite{Koscheev-Nature-2001}. However, formation of a uniform truly single layer is notoriously difficult and it is impossible to exclude a possibility that an ion penetrates the ND particle in a top layer and is implanted with smaller energy into underlying grain. In addition, presence of doubly charges ions during the implantation experiment is not fully excluded.

In this work, we use the molecular dynamics (MD) approach and apply it to the
question of the single and bimodal temperature-release of NGs. To the best of
our knowledge, atomistic simulation methods have not previously been applied to
this problem. We simulate the implantation and pyrolysis processes for He and
Xe, and find excellent agreement with experimental data collected from NDs. We
reproduce the single release peak for He and bimodal release for Xe. The
simulations explain how the mass, implantation depth and crystallographic
location of the NGs gives rise to the experimental observations, and for the
first time provide an atomistic explanation of the release mechanism.
Although we do not provide an explanation for the origin of
isotopic anomalies in NDs, our results are useful in the estimation of the
composition of end-members of P3 and HL components and in studies of isotopic
fractionation of released NGs.

Our paper is structured as follows. In the methodology we introduce our MD
approach, which includes an Arrhenius framework to map the simulation
temperatures onto their experimental equivalent. Our first results are
qualitative, using visualization and movie of the implantation and thermal
release processes to highlight the differences between He and Xe. The second
set of results involves robust statistical analysis of a large number of
simulations spanning many implantation energies, implantation directions and
annealing temperatures. The final section uses the Arrhenius approach to make
direct comparison between meteoritic data and the simulations.

\section{Methodology}
\label{Meth}

\subsection{Simulation methods}
\label{SimMeth}

The MD simulations are performed using the Environment Dependent Interaction
Potential (EDIP) \cite{Marks-PRB-2000} for C--C interactions in combination
with the standard Ziegler-Biersack-Littmark (ZBL) potential \cite{ZBL-1985} to
describe close approaches (see \cite{aghajamali_thesis} for further details).
EDIP has proved itself to be highly transferable \cite{deTomas-Carbon-2016,
deTomas-Carbon-2019} and has previously been successful in simulations of many
different forms of C such as diamond \cite{Marks-PRL-2012, Fairchild-AM-2012,
Buchan-JAP-2015, Regan-AdvancedMaterials-2020}, carbon onions
\cite{Lau-PRB-2007}, amorphous carbon \cite{Marks-PRB-2002,
best2020evidence}, nanotubes \cite{SuarezMartinez-Carbon-2012}, peapods
\cite{SuarezMartinez-Carbon-2010} and carbide-derived carbons
\cite{deTomas-Carbon-2017, deTomas-APL-2018}. One aspect of 
EDIP which is
still being developed is the ability to describe hydrogen; accordingly, none
of the simulations presented here include hydrogen.
To describe He--C and Xe--C
interactions, we use a Lennard-Jones (LJ) potential coupled with the ZBL
potential, with interpolation between the two interactions controlled using
Fermi-type switching functions as described in \cite{Buchan-JAP-2015} and
\cite{Christie-Carbon-2015}. This combination of the LJ and ZBL potentials is
identical to our recent article \cite{Fogg-NIMB-2019} where we studied the
modification of NDs by Xe implantation. Full details are provided in the
Supplementary Information.

We perform our simulations using an in-house MD package. Implantation
simulations are carried out in an NVE ensemble (meaning number of particles
(N), volume (V) and energy (E) are conserved quantities), using Verlet
integration and a variable timestep \cite{Marks-NIMB-2015}. Annealing
simulations are performed in an NVT ensemble using a velocity-rescaling thermostat.
Periodic boundary conditions are not employed. All simulations use a 4999-atom
ND with a diameter of 3.9~nm; this size is similar to synthetic and meteoritic
NDs \cite{Huss-Met-1994-I, Huss-Met-1994-II, Koscheev-Nature-2001} (meteoritic
NDs are characterized by a log-normal size distribution with a median of 2.6~nm
and maximal sizes up to 10~nm or more \cite{Lewis-Nature-1987}). 

\begin{figure} [t]
\centering
\includegraphics[width=1\columnwidth]{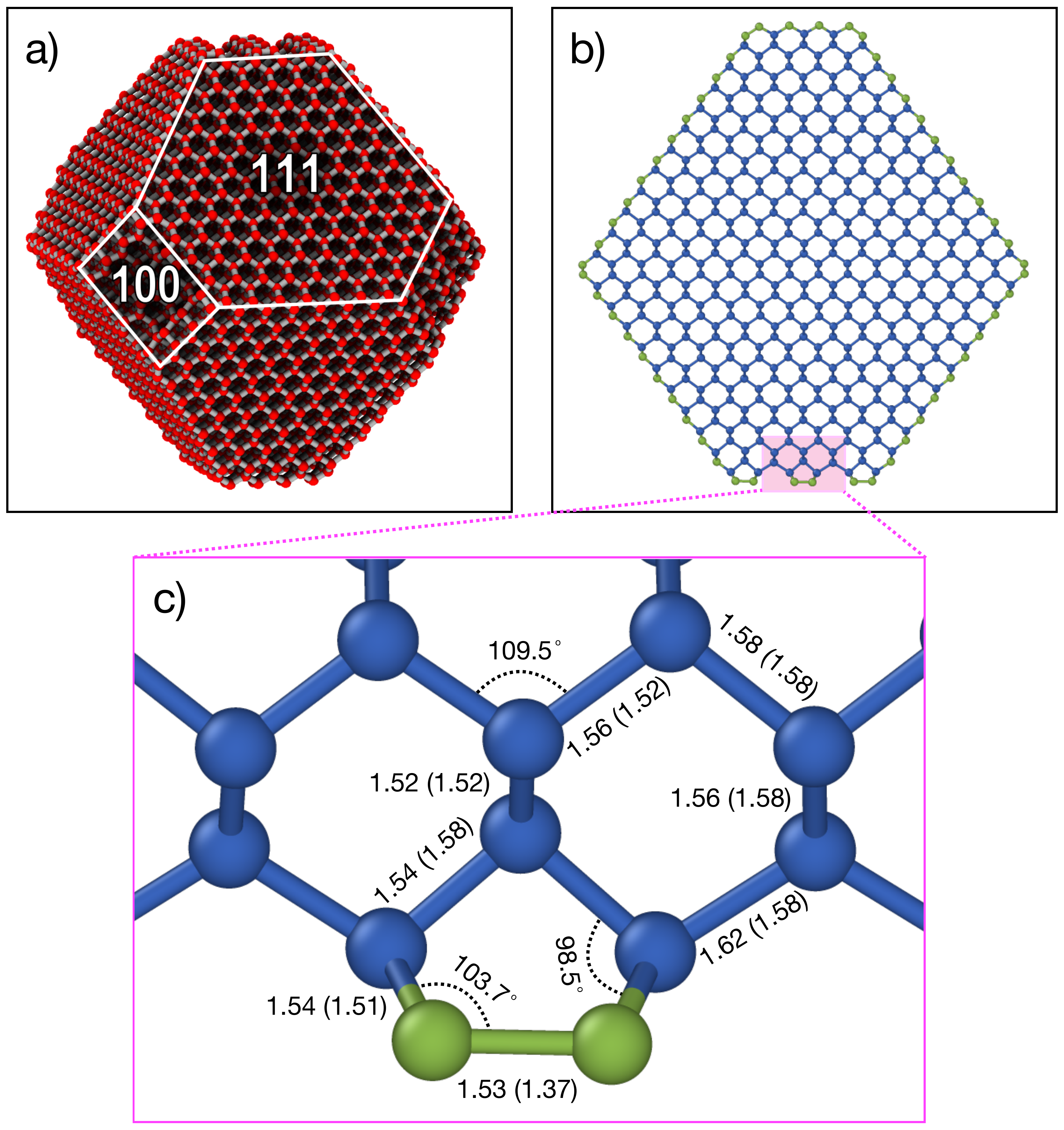}
\caption{ Views of the 3.9~nm diameter ND used in this work. (a) Perspective
view showing the truncated octahedral geometry. C atoms are shown in red and
the white lines highlight the \{100\} and \{111\} faces. (b) Cross-sectional
view (1~nm slice) using color coding to indicate the hybridization: $sp^2$--C
and $sp^3$--C atoms are shown as green and blue circles, respectively. Note
that the \{100\} faces are reconstructed in a 2$\times$1 manner to eliminate
dangling bonds. (c) Magnified view of the shaded pink region in
panel (b), highlighting the local environment in the reconstructed face of the
ND. Bond lengths are in \AA\ and density-functional-theory values from de La Pierre, \emph{et al.} \cite{de2014100} are shown in parentheses.  }
\label{Figure1}
\end{figure}

As shown in Figure~\ref{Figure1}, the ND has a truncated octahedral form which
is the stable geometry for dehydrogenated NDs \cite{Barnard-JCP-2004}. The
\{100\} faces of the ND are reconstructed in a 2$\times$1 manner to eliminate
dangling bonds. The local environment, bond length and bond
angles in this reconstructed geometry are shown in Figure~\ref{Figure1}(c) and
compared with density-functional-theory (DFT) data from de La Pierre, \emph{et al.} \cite{de2014100}. Most
of the bond lengths are very similar (within 0.04~\AA), the only exception
being the $sp^2$--$sp^2$ bond which is 0.16~\AA\ larger with EDIP.  The \{111\}
faces are not reconstructed, as this is the stable configuration with EDIP. We
note that EDIP does not predict the 2$\times$1 $\pi$-bonded chain
reconstruction \cite{pandey1982new} of the \{111\} face; EDIP favours the
1$\times$1 geometry by 3.6~J/m$^2$, whereas DFT calculations \cite{de2014100}
favour the 2$\times$1 geometry by 0.93~J/m$^2$. We do not expect the 1$\times$1
geometry preferred by EDIP impacts on our results because both reconstructions
graphitize in a smilar manner and produce the same final structure and
orientation relative to the rest of the ND.

\begin{figure*}[!t]
\centering
\includegraphics[width=0.86\textwidth]{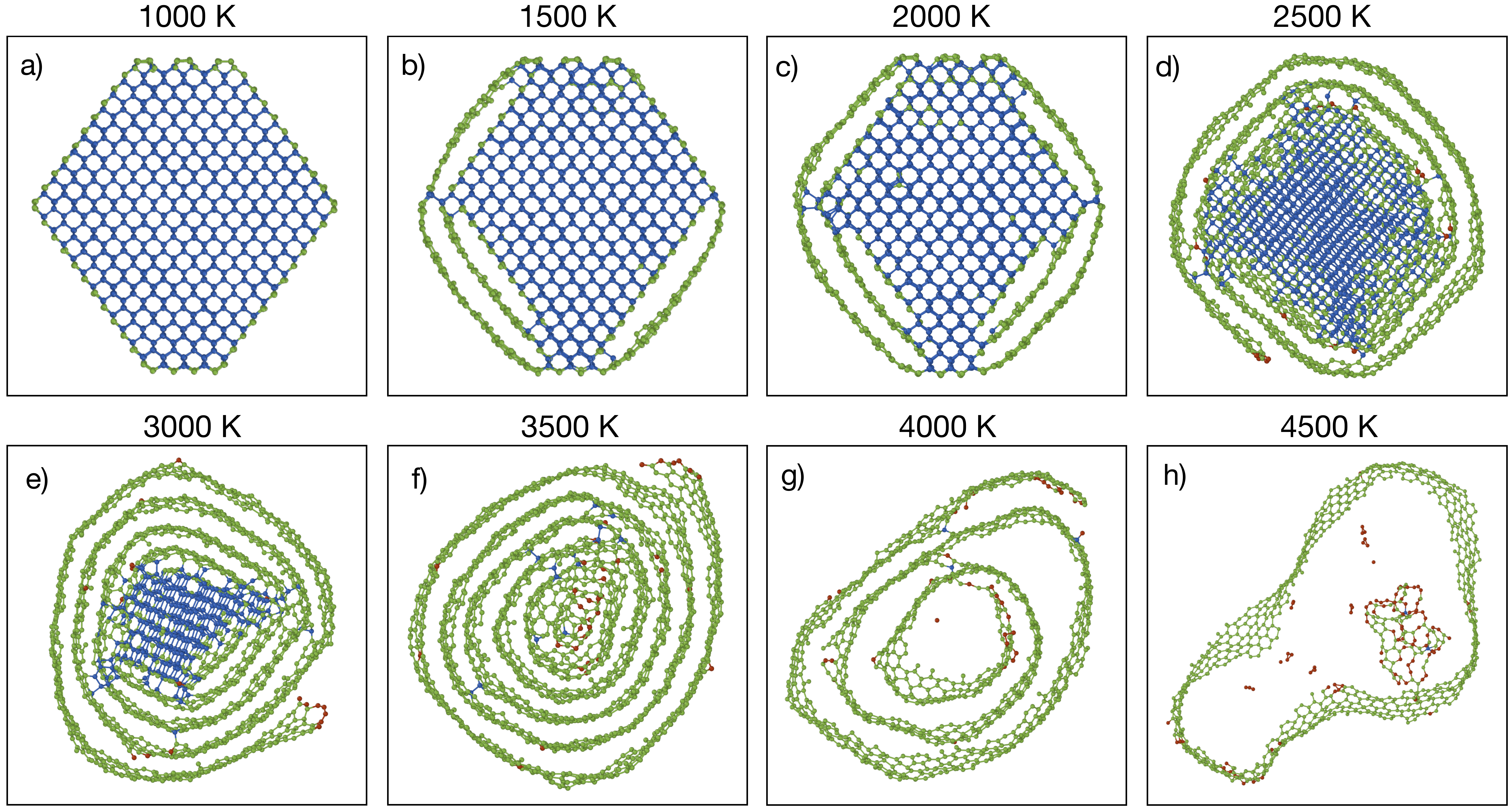}
\caption{ Cross-sectional snapshots (1~nm slices) of pristine ND after annealing for 1~ns at various temperatures. Panel (f) shows that at 3500~K the experimentally observed result, namely a carbon onion, is obtained. Red, green and blue circles denote $sp$--C, $sp^2$--C and $sp^3$--C hybridization, respectively. Full discussion of the onionization process is provided in the Supplementary Information. }
\label{Figure2}
\end{figure*}

The coordinates of the ND were generated using a methodology described in our
recent article studying Xe implantation \cite{Fogg-NIMB-2019}. The key idea is
to cut the ND out of an infinite crystal using clipping planes in the
$\langle$100$\rangle$ and $\langle$111$\rangle$ directions. Using the notation
of Fogg, \emph{et al.} \cite{Fogg-NIMB-2019}, the ND used in this work was generated with
$d_{100}$=20~\AA\ and $d_{111}$=17~\AA. Prior to implantation, the ND is
equilibrated at 300~K.

Implantation simulations are 1~ps in length, sufficient to model the ballistic phase of ion implantation into the ND. To generate a wide variety of implantation conditions, the initial position and direction of the implanted species (either He or Xe) was systematically varied. Following our previous studies \cite{Buchan-JAP-2015, Christie-Carbon-2015, Shiryaev-SciRep-2018, Fogg-NIMB-2019}, the initial position is taken from a 25-point solution to the Thomson problem \cite{Thomson-PM-1906} which distributes coordinates uniformly on a sphere (the Cartesian coordinates are provided in \cite{aghajamali_thesis}). Some implantations are performed directly towards the centre of the ND, while others are directed slightly (up to 10~\AA) away from the centre-of-mass. After the system equilibrates, the implantation depth is computed relative to the nearest crystallographic face, \{100\} or \{111\}. For the He implantations the mass was 4~amu, while the Xe implantations use a mass of 133~amu, the same as in \cite{Shiryaev-SciRep-2018, Fogg-NIMB-2019} and slightly higher than the average isotopic value of 131.3~amu. Additional Xe simulations use masses of 124 and 136~amu, corresponding to the lightest and heaviest stable isotopes. 

Annealing simulations extending up to 1~ns are performed to study the thermal release process. Coordination analysis is performed by counting the number of nearest neighbours within a cutoff of 1.85~\AA. For the purposes of analysis and visualization, C atoms are considered to be $sp$, $sp^2$ and $sp^3$ hybridized if they have two, three and four neighbours, respectively. Visualization is performed using the \textsc{OVITO} package \cite{Stukowski-MSMSE-2010}. 

\subsection{Mapping to experimental temperatures}
\label{Arrhenius-formula}

Generally speaking, the timescale of MD simulations are on the order of nanoseconds, around 13 orders shorter than the experimental annealing time. This enormous difference complicates comparison between simulations and experiments, as thermally activated events will be suppressed in the simulations due to vastly shorter time. Many different solutions to the MD timescale problem have been proposed \cite{Voter-PRL-1997,Voter-PRB-1998,So/rensen-JChemPhys-2000}, but here we employ a simple temperature-acceleration approach that we have used successfully on other C systems \cite{deTomas-Carbon-2017}. The first step is to assume Arrhenius behavior and a single activation energy, which corresponds to the relation 
\begin{equation} f=A~\mathrm{exp}(- E_a/k_B T) \end{equation} 
where $f$ is the frequency of events, $A$ is the attempt frequency, $T$ is the temperature, $k_B$ is Boltzmann's constant and $E_a$ is the activation energy. The correspondence between the experimental and simulation temperature is determined by equating the time-frequency product (i.e., $f_\mathrm{expt}\times t_\mathrm{expt}=f_\mathrm{sim}\times t_\mathrm{sim}$) to ensure that for the same activation energy the same number of events occur in both simulation and experiment. This yields the following expression 
\begin{equation} T_\mathrm{sim} = - \frac{E_a}{k_B} \times \left[ \log \left(
\frac{t_\mathrm{expt}}{t_\mathrm{sim}} \right) 
			- \frac{E_a}{k_B T_\mathrm{expt}} \right]^{-1}
			  \label{equ2} \end{equation} 

\begin{figure} [!b]
\centering
\includegraphics[width=1\columnwidth]{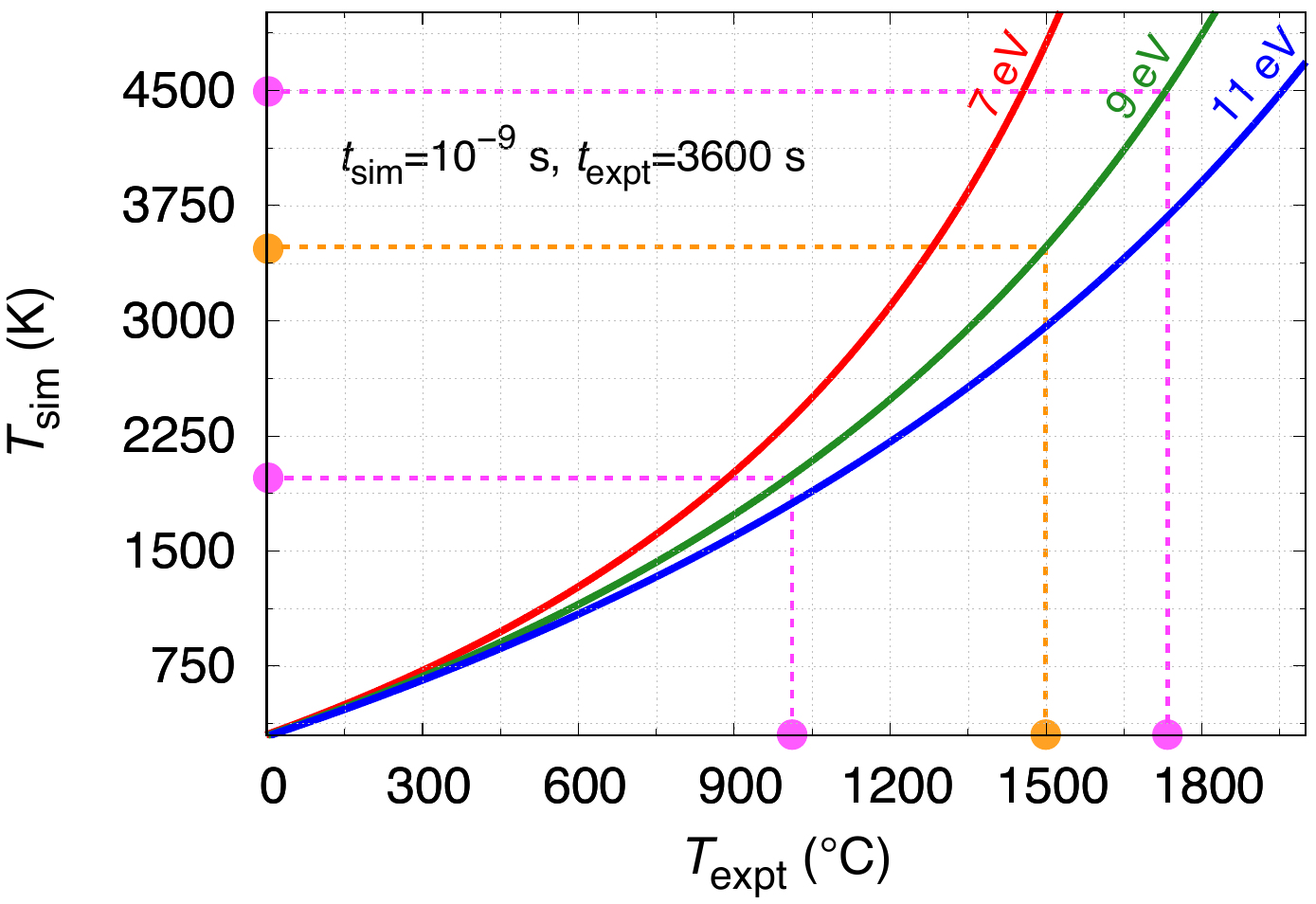}
\caption{ Calibration curve between simulation and experimental temperatures
via the Arrhenius approach. The orange dashed line indicates data for the 
onionization of a ND used to determine the value of $E_a$=9~eV. The pink lines 
indicate two examples of the mapping process as described in the text. }
\label{Figure3}
\end{figure}

This equation links the experimental temperature and time ($T_\mathrm{expt}$ and $t_\mathrm{expt}$) with those of the simulation ($T_\mathrm{sim}$ and $t_\mathrm{sim}$) with the only parameter being the activation energy. To determine $E_a$, we make use of experimental data on graphitization of NDs, where it is known that $\sim$1500$^\circ$C is required to convert NDs into a carbon onion \cite{Kuznetsov-Carbon-1994,Kuznetsov-ChemPhysLett-1994,Tomita-Carbon-2002, Qiao-ScriMat-2006,Xiao-NanoLett-2014}. To identify the correponding temperature on the MD timescale we performed a set of 1~ns simulations at different annealing temperatures and found that onionization of the ND occurs at around 3500~K (see Figure~\ref{Figure2}). Using these two temperatures and suitable times ($t_\mathrm{sim}$=$10^{-9}$~s and $t_\mathrm{expt}$=3600~s \cite{Kuznetsov-ChemPhysLett-1994,Kuznetsov-Carbon-1994,Tomita-Carbon-2002, Qiao-ScriMat-2006,Xiao-NanoLett-2014}), we obtain via Equation~\ref{equ2} a value of $E_a$=9~eV. Having determined the activation energy, we can employ Equation~\ref{equ2} to map any simulation temperature to its experimental equivalent using the relation shown in Figure~\ref{Figure3}. For example, the pink lines in the figure show that a simulation temperature of 4500~K is equivalent to an experimental temperature of $\sim$1730$^\circ$C, and a simulation at 2000~K is equivalent to an experiment at $\sim$1000$^\circ$C. The latter also serves as a convenient dividing line between main peaks of Xe-P3 and Xe-HL components \cite{Huss-Met-1994-I,Huss-Met-1994-II}. To provide a sense of scale, Figure~\ref{Figure3} also shows calibration curves for two other activation energies. Note that due to the logarithm term in Equation~\ref{equ2} the value of $E_a$ has minimal sensitivity to the choice of $t_\mathrm{expt}$. This point is discussed in detail in de Tomas, \emph{et al.} \cite{deTomas-Carbon-2017}. Finally, we note that throughout this manuscript we adopt the convention that experimental temperatures are given in Celsius and simulation temperatures are in Kelvin. This helps conceptually separate the two quantities which otherwise might be confused with one another.

\section{Results and Discussion}
\label{Results}

\subsection{Individual implantation and thermal release events}
\label{A}

\begin{figure}[!t]
\centering
\includegraphics[width=1\columnwidth]{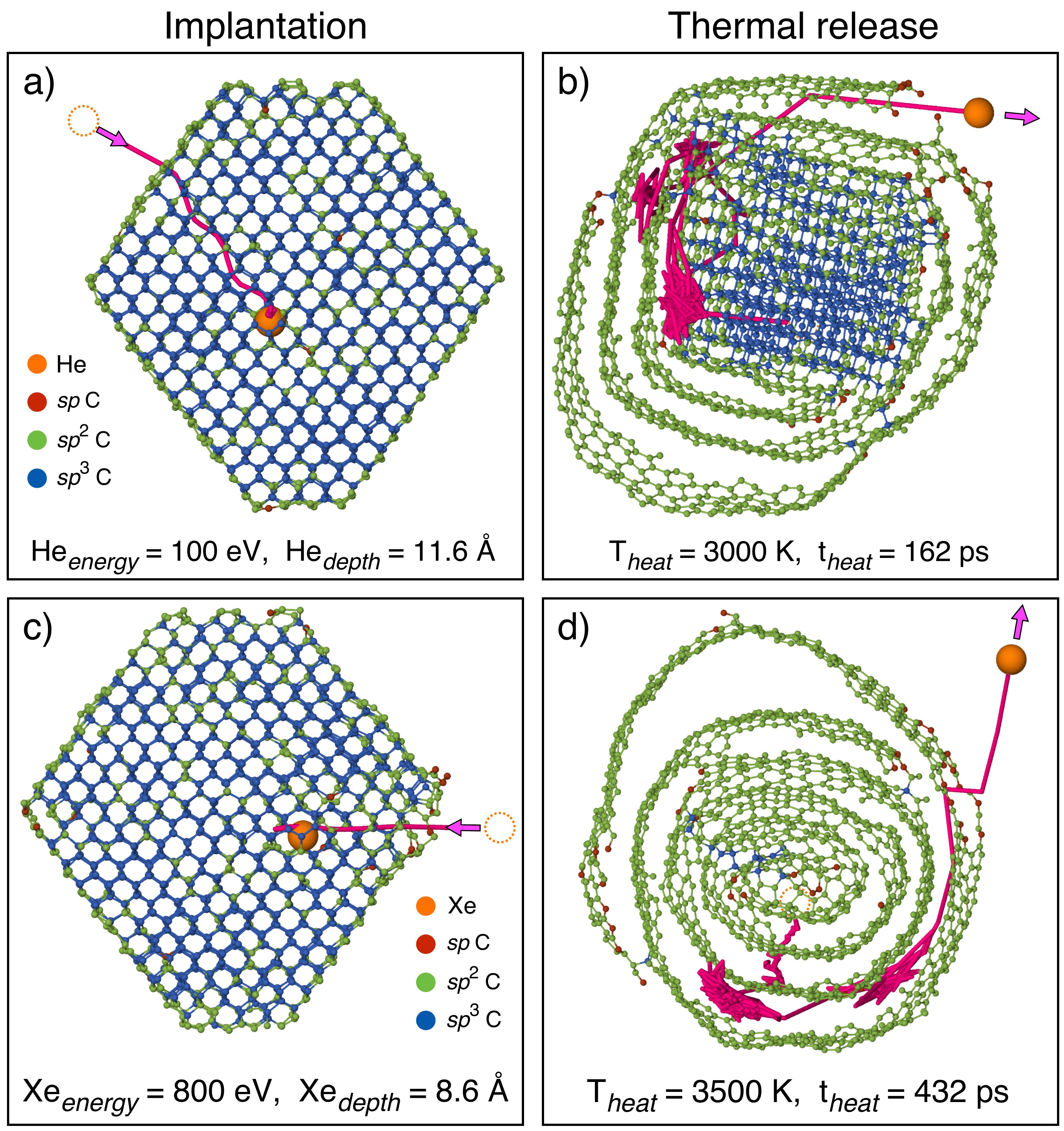}
\caption{ Typical implantation (left panels) and thermal release (right panels)
processes involving He [panels (a) and (b)] and Xe [panels (c) and (d)]. The
pink lines indicate the He and Xe trajectories, and the He and Xe
are shown as an orange circles. The color codings and cross-sectional slice
details are the same as Figure~\ref{Figure2}.}
\label{Figure4}
\end{figure}

Representative examples of the implantation and thermal release processes for He and Xe are shown in Figure~\ref{Figure4}. Panels (a) and (c) show that the implantation processes for He and Xe differ substantially, with the higher mass and size of the Xe having a large effect. For He, only a relatively modest energy ($\sim$100~eV) is needed to implant the atom into the centre of the ND, and the He undergoes multiple deviations along the implantation trajectory (pink line) since the C atoms are three times heavier. In contrast, the implanting Xe simply slows down, and these are the C atoms which move. Around $\sim$800~eV of kinetic energy is needed to implant into the central region, similar to the value of circa 700~eV used by Koscheev, \emph{et al.} \cite{Koscheev-Nature-2001} in their experiments. 

Once the implanted ND system has equilibrated, the entire cluster is heated to find the temperature at which the NG atom escapes. Examples of this process are shown in panels (b) and (d). For both species, thermal release occurs at relatively high simulation temperatures; 3000~K for He and 3500~K for Xe. As seen in Figure~\ref{Figure2}, these temperatures are sufficient to transform portions of the ND into a onion-like structure. Due to its small size, the He is able to escape before the ND has completely onionized. The complexity of the process can be appreciated in the first part of the Supplementary Movie which shows how the He travels along multiple $\langle$110$\rangle$ channels, and traverses a substantial fraction of the ND before escaping. In the case of Xe, the ND is fully transformed into a carbon onion and the release process is more difficult as the Xe must pass through the graphitic shells. The pink trajectory in panel (d) shows that the release process involves two local minima in which the Xe jiggles back-and-forth many times before escaping. An animation sequence of this process is provided in the second part of the Supplementary Movie. 

\begin{figure} [!t]
\centering
\includegraphics[width=1\columnwidth]{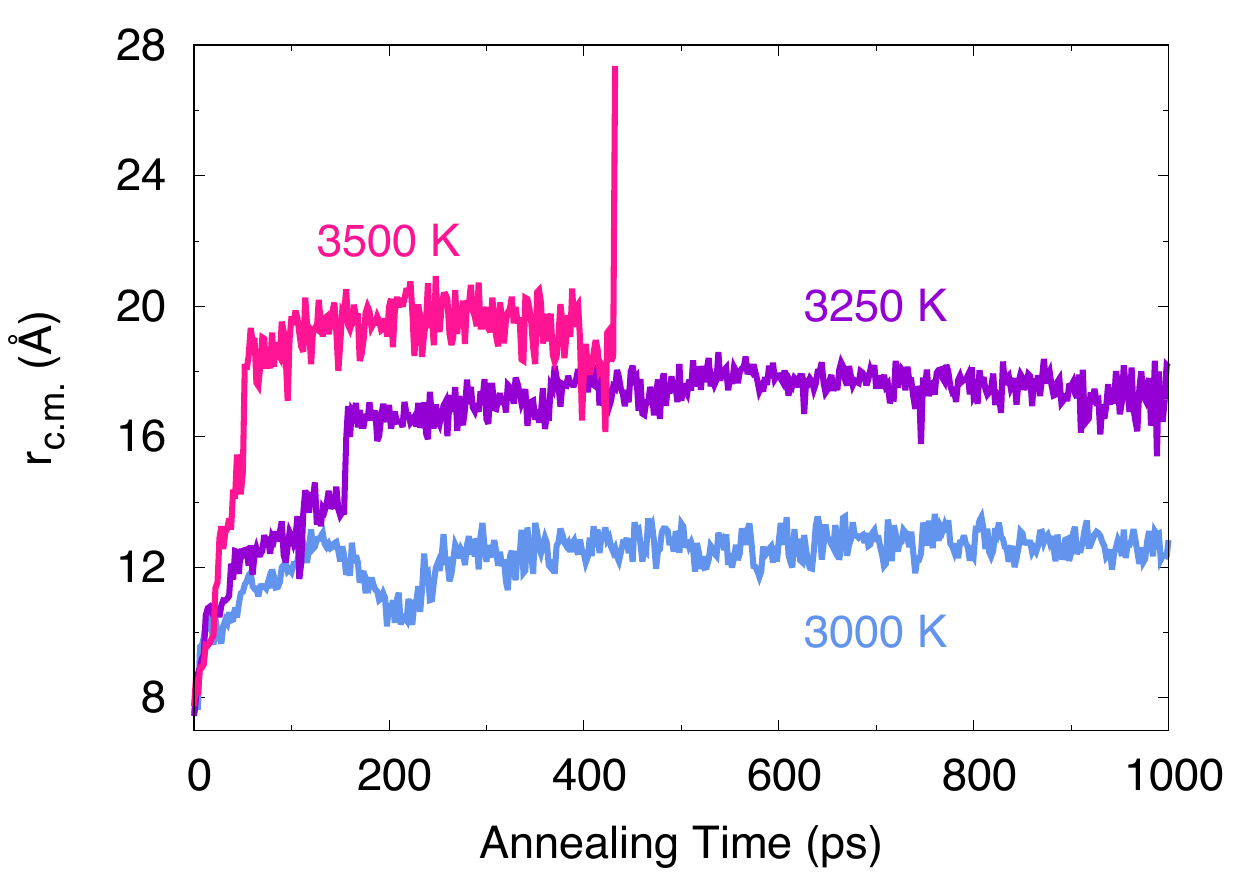}
\caption{Distance between Xe and the center of mass of the ND as a function of annealing time for three different annealing temperatures. The structure in Figure~\ref{Figure4}(c) is the starting point for all three simulations. }
\label{Figure5}
\end{figure} 

To determine the temperature at which He and Xe are released we perform 1~ns annealing simulations at many different temperatures and monitor the distance between the NG species and the center of the mass of the system. An example of our methodology is provided in Figure~\ref{Figure5} which shows this quantity ($r_\mathrm{c.m.}$) for Xe as a function of time for three different annealing temperatures, using the implanted ND in Figure~\ref{Figure4}(c) as the starting structure. At 3000~K (blue line) the Xe remains the same distance from the center of mass. This occurs because the Xe becomes trapped at the interface between the ND core and onionized outer layers (see Figure~\ref{Figure2}(e)). At 3250~K (violet line) the Xe moves towards the surface of the ND during the first $\sim$170~ps of annealing, but afterwards it is trapped between the graphitic sheets. Note that at around 30 and 160~ps, there are two significant surges in $r_\mathrm{c.m.}$ where the Xe atom passes through the graphitic sheets. At 3500~K (pink line), the Xe is initially trapped between the sheets, but after moving between them, finally exits through a gap in the outer shell at 432~ps, producing a sharp jump in $r_\mathrm{c.m.}$.

\subsection{Statistical analysis}
\label{B}

\begin{figure} [!b]
\centering
\includegraphics[width=1\columnwidth]{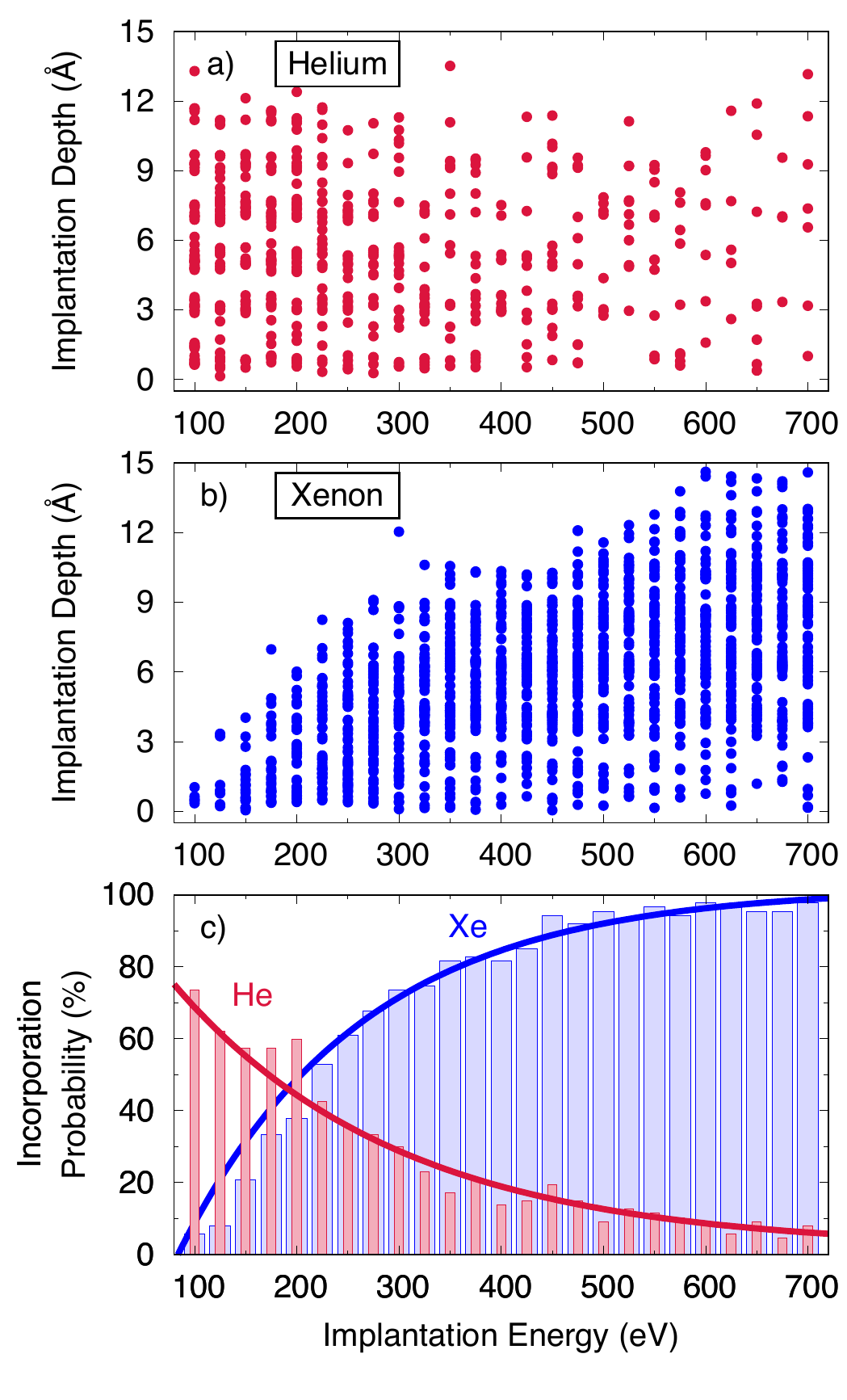}
\caption{Implantation depth of (a) $^{4}$He and (b) $^{133}$Xe as a function  of implantation energy. Panel (c) shows the incorporation probability of  implanted He (red) and Xe (blue) as a function of implantation energy. The solid lines in (c) are exponential fits to guide the eye, and have decay  constants of 0.22 and 0.16~eV for He and Xe, respectively. }
\label{Figure6}
\end{figure}

To collect a statistically significant data set a large number of implantation and thermal release simulations need to be performed. The first step is to perform simulations that implant He and Xe into a wide variety of configurations within the ND.  For each species a total of 1875 implantations were performed; 25 different energies, each with 75 different initial conditions (i.e.,\ directions and/or impact parameters). A summary of the resultant implantation depth and incorporation probability is shown in Figure~\ref{Figure6}. Panels (a) and (b) show the implantation depth of He and Xe as a function of implantation energy, where each dot indicates an implantation event whereby the NG species remains with the ND. Impacts where the He or Xe leave the ND are not shown. The fewest dots occur for high He implantation energies (where He passes through the ND) and low Xe implantation energies (where Xe is reflected). The distribution of depths differs considerably between the two species. For He, the implantation depth spans the maximum possible range, extending from the surface to the centre of the ND, and the implantation depth is uncorrelated with energy. In contrast, low energies result in only shallow implantation of Xe, while nearly 600~eV is required to span the full range of depths. High Xe implantation energies still produce a large number of shallow implantation depths, which is perhaps due to the small size of the ND.

Figure~\ref{Figure6}(c) quantifies the incorporation probability of He and Xe as a function of implantation energy. The probabilities for the two species are strikingly different, and are driven by the mass difference relative to C. This data shows that low energies are optimal for He implantation, while efficient Xe implantation requires many hundreds of eV. Noting that He is the most abundant NG in meteoritic NDs \cite{Huss-Elements-2005}, this data imposes constraints on the astrophysical conditions for He incorporation via implantation. 

\begin{figure}[!t]
\centering
\includegraphics[width=1\columnwidth]{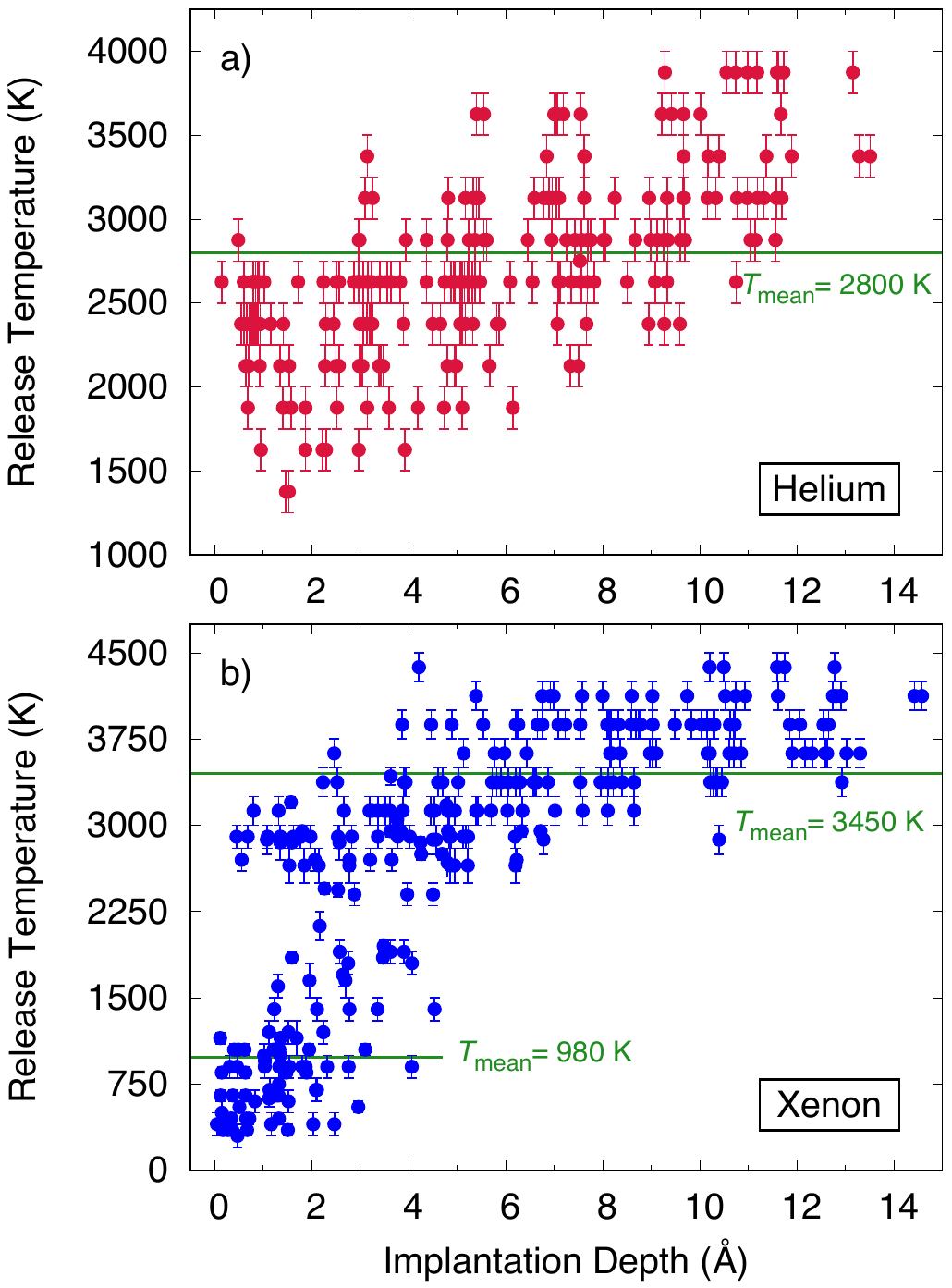}
\caption{Thermal release data of (a) $^{4}$He and (b) $^{133}$Xe as a function of implantation depth. Mean release temperatures ($T_\mathrm{mean}$) are shown by horizontal green lines and error bars indicate the degree of uncertainty; the precise meanings are explained in the text. }
\label{Figure7}
\end{figure}

The second step in performing the statistical analysis uses the coordinates associated with each dot in Figure~\ref{Figure6} as the starting structure of thousands of thermal release simulations. All simulations run for 1~ns, except for when $r_\mathrm{c.m.}$ indicate that release has occurred; in such cases the simulation is terminated. If thermal release does not occur, the simulation is rerun at a higher temperature. Typically, the temperature increment between successive simulations is 100--250~K. The raw simulation data showing the relationship between the release temperature of the NG and the implantation depth is shown in Figure~\ref{Figure7}. The dots and error bars indicate the precision, meaning that if a NG atom releases at a temperature $T_2$ but not at a lower temperature $T_1$, then the dot denotes $(T_1+T_2)/2$ and the error bar indicates the range $[T_1:T_2]$. Panel (b) shows that the Xe release temperatures cluster into two groups and correlate with the depth of the implanted atom. The average temperature for these two clusters are indicated by the green lines, with a dividing line of 2000~K used to separate the groups. Varying this number by several hundred Kelvin makes little difference to the averages. For He, no clustering occurs, and the release temperature gradually increases with implantation depth. In this case the green line indicates the average temperature for the full data set.

\subsection{Comparison with experiment}

\begin{figure}[!b]
\centering
\includegraphics[width=1\columnwidth]{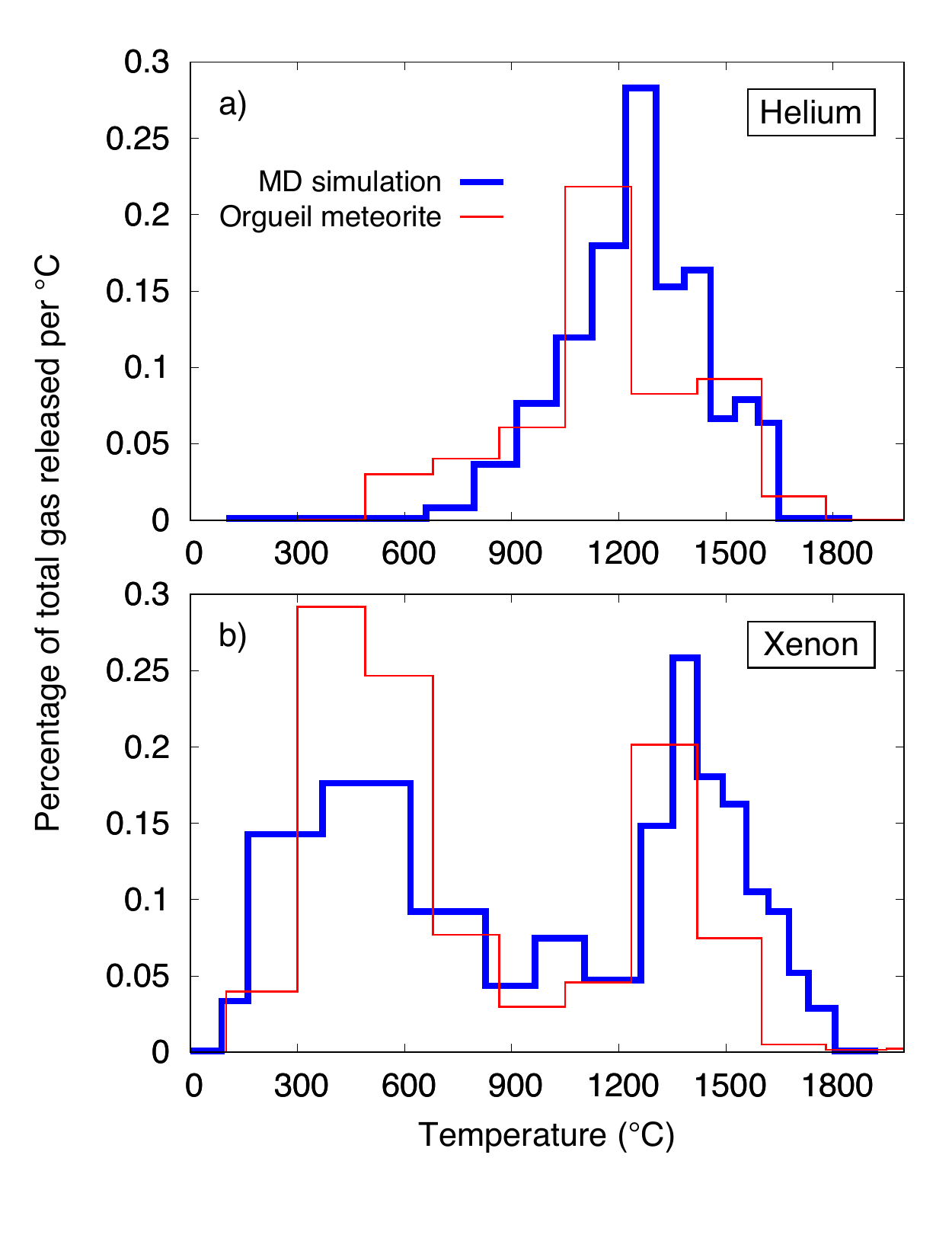}
\caption{ Measured and simulated ND thermal release patterns for (a) $^{4}$He and (b) $^{133}$Xe. Experimental values (red lines) are taken from data from the Orgueil meteorite ND \cite{Huss-Met-1994-I,Huss-Met-1994-II}, while the MD values (blue lines) are from this work. }
\label{Figure8}
\end{figure}

Figure~\ref{Figure7}(a) shows that a minimum of 1300~K is required to release He and by 4000~K all He is released. In contrast, the Xe data in panel (b) spans a broader range, with a minimum release temperature of only 400~K and a maximum of nearly 4500~K, at which point the ND is effectively destroyed (see Figure.~\ref{Figure2}(h)). The observation of two temperature clusters for Xe and a single broad distribution for He is in good qualitative agreement with the meteoritic ND observations. To map the simulation temperatures onto their experimental equivalents, we employ the Arrhenius approach explained in the Methodology, using an activation energy of $E_a$=9~eV. By histogram binning the simulation data in Figure~\ref{Figure7} and applying the transformation in Equation~\ref{equ2} we obtain a data set shown as thick blue lines in Figure~\ref{Figure8}. On the same scale we show experimental data for the Orgueil meteoritic NDs extracted from Huss and Lewis \cite{Huss-Met-1994-I, Huss-Met-1994-II}. The agreement is remarkable, with the simulations reproducing all of the main meteoritic ND characteristics, including (i) the unimodal vs bimodal character, (ii) the position of the peaks, (iii) the widths of the distributions, and (iv) the maximum and minimum release temperatures. This is the first time that MD simulation has predicted these important effects.

The high level of experimental detail reproduced by the simulations provides post-hoc justification for the Arrhenius approach, confirming that the presumption of a dominant activation energy is reasonable for this class of problem. All of the predicted temperatures are correctly positioned relative to their experimental equivalents; this includes the onset of He and Xe release, the position of the release peaks, and even the upper limit at $\sim$1800$^\circ$C, which corresponds to destruction of the ND and the release of any remaining gases. Regarding Figure~\ref{Figure8}(b), it is important to note that the simulations cannot predict the relative height of the peaks in the bimodal distribution, since this is function of the balance between shallow and deeply implanted Xe which is not known.

\begin{figure}[b]
\centering
\includegraphics[width=1\columnwidth]{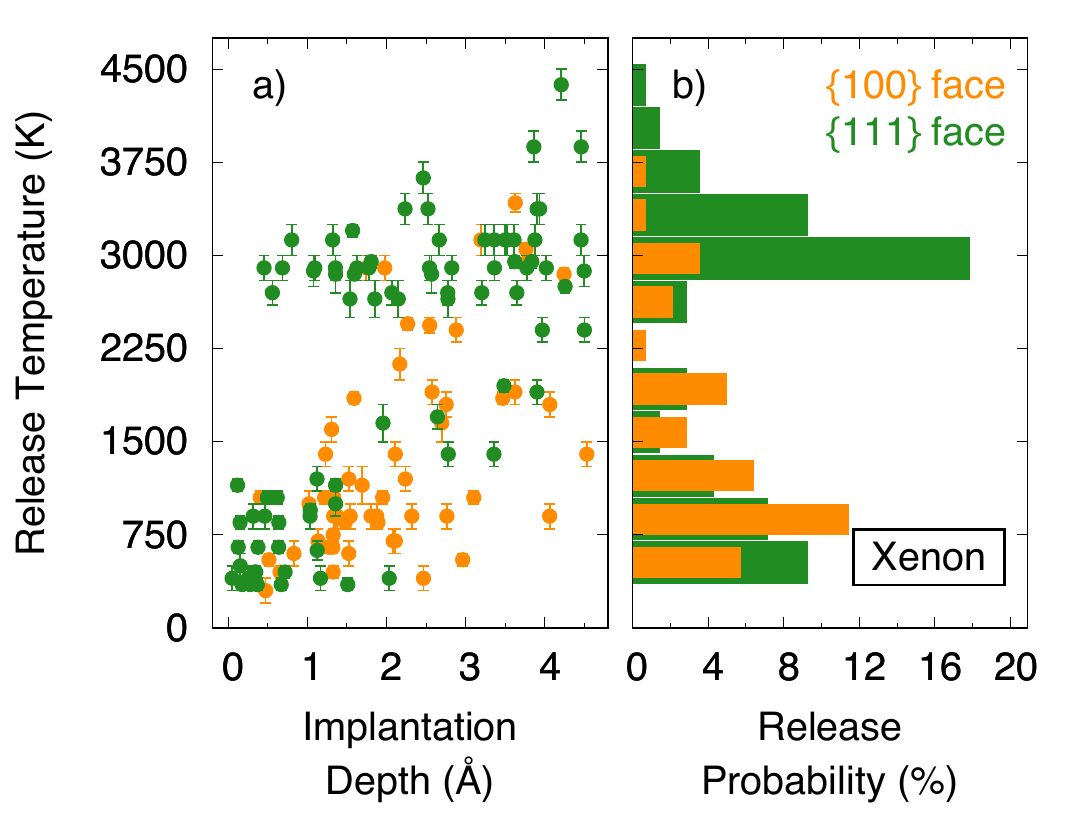}
\caption{ (a) Classification of shallow-implantation Xe data in Figure~\ref{Figure7} according to proximity to a \{100\} faces (orange circles) or a \{111\} faces (green circles). (b) Histogram analysis of the data in (a). }
\label{Figure9}
\end{figure}

Having reproduced the essential characteristics of NG release, we can address some of the most fundamental questions related to meteoritic NDs. Specifically, we can examine the atomistic origin of the low- and high-temperature peaks of Xe release, and understand why He exhibits unimodal behaviour. Considering first the question of Xe, the raw simulation data in Figure~\ref{Figure7}(b) provides clues as to why release occurs at two distinct temperatures. The low-temperature peak is seen to be strongly correlated with proximity to the surface, with 4.5~\AA\ being the critical depth beyond which low-temperature release is impossible. In contrast, high-temperature release is possible for all implantation depths, and even for depths around 1~\AA\ there are two distinct temperature-release populations. The origin of this behaviour is that the surface of the ND progressively graphitizes with temperature, with the \{111\} faces transforming prior to the \{100\} faces (see Figure~\ref{Figure2}). As a result, Xe located close to a \{111\} faces can become trapped by the developing graphitic layers, and once the layer has fully formed, the Xe is too large to easily diffuse through the hexagonal graphene-like network. This effect is quantified in Figure~\ref{Figure9} which replots the Xe data in Figure~\ref{Figure7}(b) for shallow depths. Panel (a) colour codes each configuration according to the closest crystallographic face, while panel (b) shows a histogram of the thermal release temperatures. It is apparent that the low-temperature peak contains roughly equal contributions from Xe near the \{100\} and \{111\} faces, while the high-temperature component is dominated by Xe close to the \{111\} faces.  Placing these observations in a gas release context, we can assert that high temperature peak (such as main peak of Xe-HL) is associated with either deeply buried Xe or shallow burial near a \{111\} faces, while the ions released at low temperatures (in particular, e.g., Xe-P3) sits just a few {\AA}ngstr\"{o}ms from the surface, and has no crystallographic preference.

\begin{figure} [!t]
\centering
\includegraphics[width=1\columnwidth]{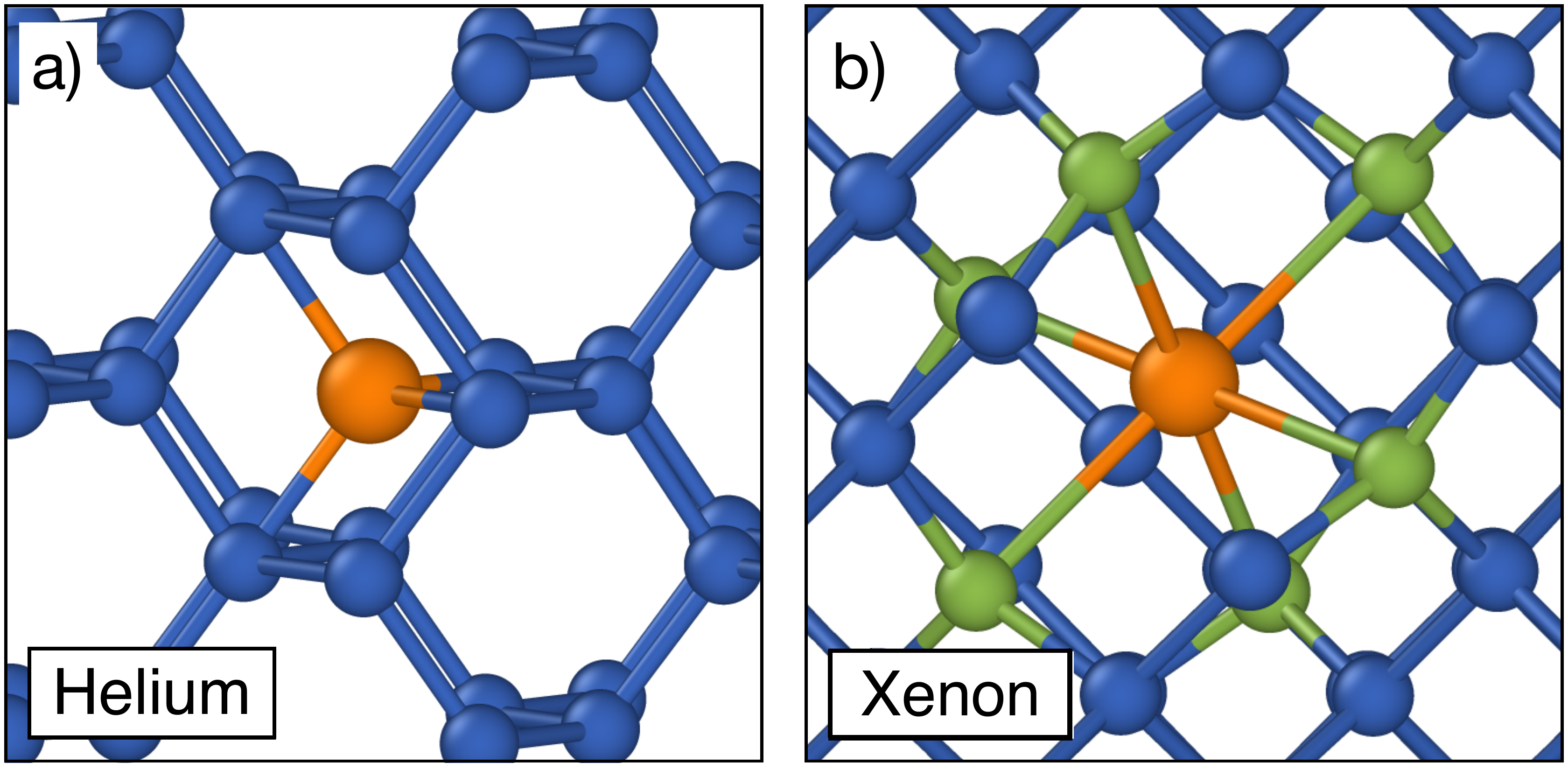}
\caption{ (a) Tetrahedral interstitial (T-site) of He and (b) Xe$_s$-V defect of Xe in ND. Colour coding details are the same as Figure~\ref{Figure4};  He and Xe are shown as orange circles. }
\label{Figure10}
\end{figure}

The other major result in Figure~\ref{Figure7} is the prediction of the
unimodal release distribution for He. The animation sequences reveal that this
behaviour can be linked to the nature of the He defect within the ND.  In
particular, the simulations show that He prefers a tetrahedral interstitial
(T-site) in the diamond lattice. Manual inspection of a large
number of implantation events show the He is always in an interstitial
position; substitutional defects and other lattice damage are not observed.
The geometry of the T-site defect is shown in Figure~\ref{Figure10}(a), with
the four orange lines between He and C highlighting the tetrahedral symmetry.
The intersitial He diffuses amongst the $\langle$110$\rangle$ channels in the
lattice, passing through a hexagonal interstitial (H-site) transition state
enroute to another T-site. This behaviour means that in the low-temperature
range, the He effectively performs a random walk around the ND. Even if the
\{111\} faces has graphitized, which typically occurs at around 1500~K in the
simulations ($\sim$800$^\circ$C in the experiments), the He cannot escape
through the graphitic layer. Only once the \{100\} faces begins to graphitize
does the He atom escape. These observations elegantly explain why the onset of
He release is so much higher than Xe.  Additionally, the smaller size of He
means that it is more mobile than Xe, which in turn explains why the peak
release temperature for He is lower than that of the corresponding Xe
high-temperature peak.

The analysis of He migration in our MD simulations is supported by DFT
calculations of NG defects in bulk diamond \cite{Goss-PRB-2009} which
similarly conclude that He forms an interstitial defect on the T-site and
diffuses via the H-site. The distances predicted by EDIP
compare well with those from DFT. For the T-site, the EDIP (DFT) values are
1.61 (1.58)~\AA\ for C--He, 1.60 (1.57)~\AA\ for the three
extended C--C bonds, and 1.52 (1.50)~\AA\ for the single compressed C--C bond.
For the H-site, the EDIP and DFT bond lengths are identical to two significant
figures, 1.57~\AA\ for the C--He distance and 1.63~\AA\ for the C--C bonds.
The prediction for the relative energetics is only qualitative, with EDIP
favouring the T-site by 0.14~eV, whereas with DFT the T-site is preferred by
2.3~eV. This large difference means that He is mobile at room temperature in
the MD simulations. The observation that the simulated He release temperature
agrees well with experimental observations shows that He release is driven by
the thermally-induced modification of the ND.

For the case of Xe defects in diamond, DFT calculations by Goss, \emph{et al.} \cite{Goss-PRB-2009}
and Drumm, \emph{et al.} \cite{Drumm-PRB-2010} show that the preferred structure is a substitutional
site involving an adjacent vacancy (Xe$_s$-V), with the Xe placed at the
midpoint (Figure~\ref{Figure10}(b)). In the relaxed structure the Xe is located
equidistance from six carbon atoms; with EDIP this distance is 2.25~\AA\ as
compared to 2.15~\AA\ in the DFT calculation of Drumm, \emph{et al.} \cite{Drumm-PRB-2010}. 
Returning our attention to the simulation data in Figure~\ref{Figure8}(b), we
can assign the low-temperature Xe release peak to an Xe$_s$-V defect near the
surface. Since this defect compromises the stability of the diamond surface,
the Xe is able to escape at modest temperatures of a few hundred degrees
Celsius. 

The DFT study by Goss, \emph{et al.} \cite{Goss-PRB-2009} also studies other NGs and divides them
into two groups: (i) He and Ne which occupy the intersitial T-site and diffuse
via the H-site, and (ii) Ar, Kr and Xe which occupy substitional sites with
vacancies. This distinction between the interstitial T-site and the
substitutional-vacancy provides a plausible explanation for the meteoritic ND
data, where He and Ne have unimodal release peaks, while Ar, Kr and Xe have
bimodal distributions. To the best of our knowledge, this connection between
defect type and temperature-release behaviour has not previously been made.

\begin{figure} [!b]
\centering
\includegraphics[width=1\columnwidth]{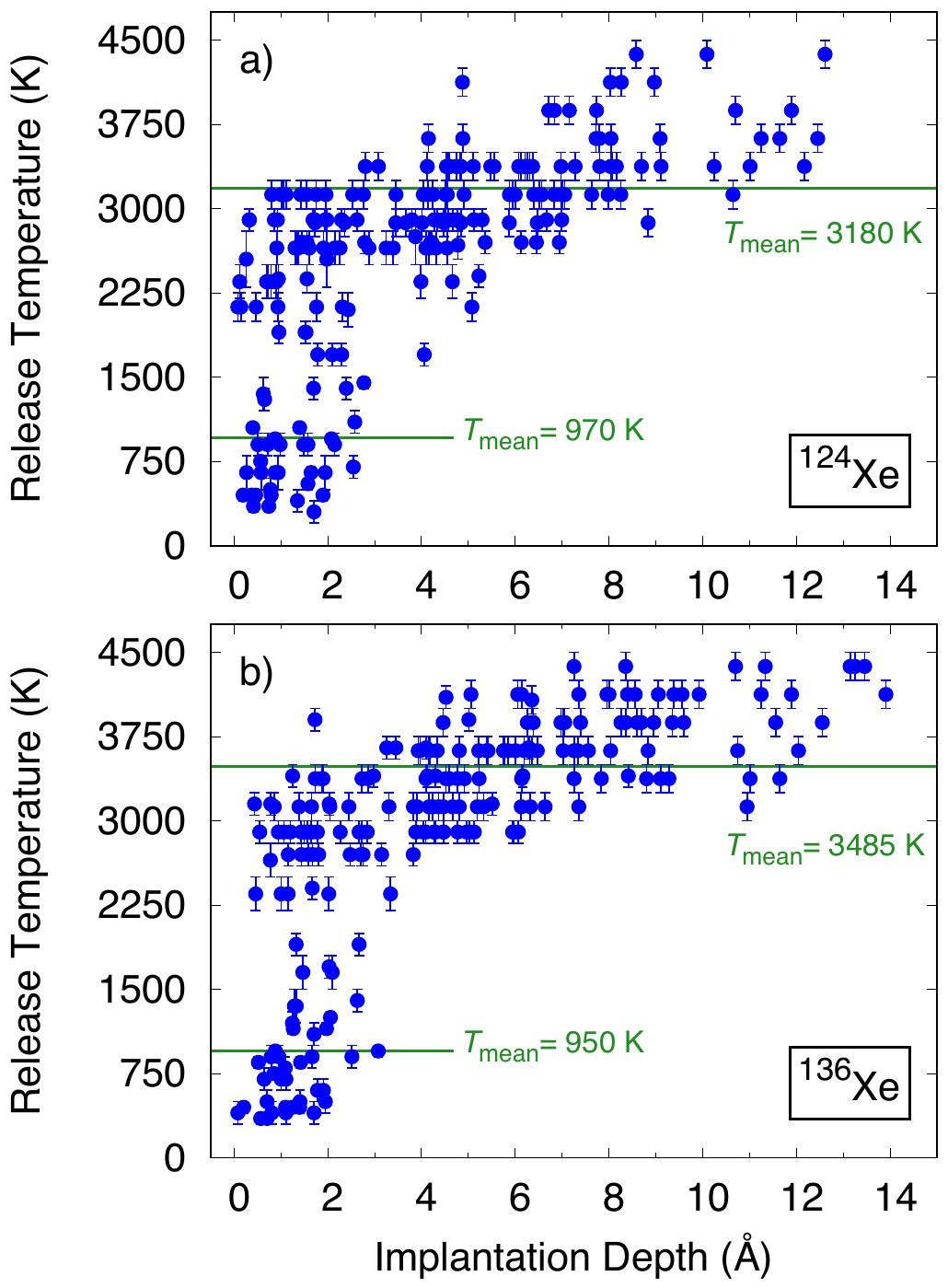}
\caption{ Thermal release data of (a) $^{124}$Xe and (b) $^{136}$Xe as a function of implantation depth. Mean release temperatures ($T_\mathrm{mean}$) are shown by horizontal green lines and error bars indicate the degree of uncertainty. Due to the computational cost, the number of data points for the two isotopes is around two-thirds of that in Figure~\ref{Figure7}(b).}
\label{Figure11}
\end{figure}

\begin{figure} [t]
\centering
\includegraphics[width=1\columnwidth]{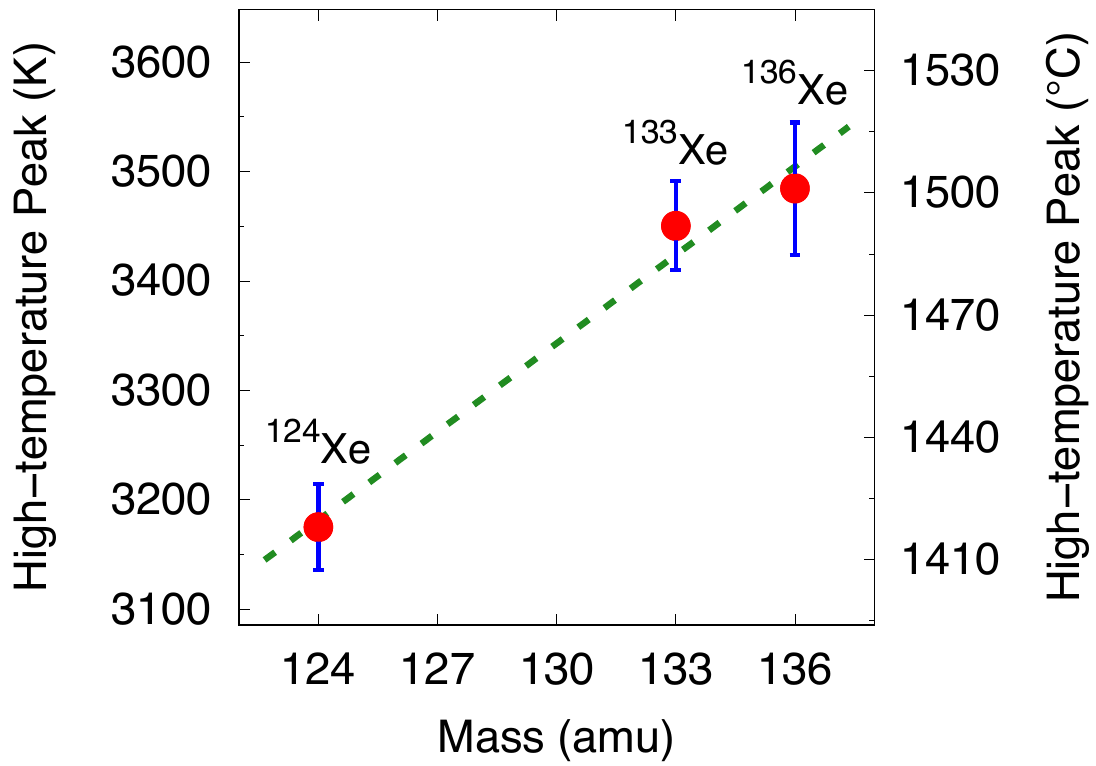}
\caption{ High-temperature release values for three different Xe isotopes as shown in Figures~\ref{Figure7}(b) and \ref{Figure11}. Left-axis  indicates raw simulation data in Kelvin while the right-axis indicates the equivalent experimental temperature in Celsius. Error bars show the standard error of the mean (SEM). The green dashed line is a linear fit to guide the eye.}
\label{Figure12}
\end{figure} 

\subsection{Isotopic effects}

Investigation of NG release kinetics from synthetic NDs reveal
marked mass-dependent fractionation for all elements
\cite{Koscheev-Nature-2001}. Namely, the gases released at high temperatures
become progressively enriched in heavy isotopes. At 1400$^\circ$C, the magnitude
of this effect reaches 0.99 \% per amu, and 1.44 \% per amu for Xe and Kr,
respectively; even larger values are observed for lighter elements
\cite{huss2008noble}. Our MD approach naturally explains this observation.

Figure~\ref{Figure11} shows thermal release data for $^{124}$Xe and $^{136}$Xe. As seen earlier, the release profile is again bimodal. For the low-temperature peak, the release temperature is similar across both isotopes, spanning a narrow range between 950 and 980~K, but for the high-temperature peak there is a large isotopic effect with the simulation release temperature varying from 3180~K for $^{124}$Xe to 3485~K for $^{136}$Xe. Figure~\ref{Figure12} plots the mean release temperature for the high-temperature peak for three Xe isotopes, with the left axis showing the simulation temperature and the right axis the equivalent experimental value; error bars denote the standard-error-in-the-mean. Between the lightest and heaviest isotope, the predicted temperature difference is over 80$^\circ$C.

The isotopic effect in Figure~\ref{Figure12} can be plausibly attributed to the effect of mass on the vibrational frequency during thermal release. An alternative explanation focusing on implantation is less attractive, since a spectrum of implantation energies are employed and there is no obvious reason why a mass difference would generate different types of defects. Regarding the thermal release process, Figure~\ref{Figure4}(d) (and Supplementary Movie) shows that the $^{133}$Xe atom escapes after an extended period of constant ``jiggling''. For about 350~ps (between $t=50$ and $t=400$~ps), the Xe is trapped between two graphitic shells and can be seen to move back-and-forth many times before eventually escaping. This observation helps explain why the lighter Xe isotopes release at the lower temperatures as the smaller mass implies a higher vibrational frequency and hence faster reaction rate.

\section{Discussion and implications to meteoritic NDs}

All calculations performed in our work are relevant for a ND grain with well-defined morphology. It is shown that proximity of the implanted atom to a given crystallographic face -- \{100\} or \{111\} -- can be important for the temperature of gas release. Here an important question arises: what is the minimal size of a ND grain, which develops crystallographic faces? For meteoritic NDs the situation is not obvious, since all imaged grains appear rounded and grains found in situ in meteorites \cite{garvie2006carbonaceous} are always associated with disordered $sp^2$--C; spectroscopic data -- EELS \cite{garvie2006surface} and XANES \cite{shiryaev2011spectroscopic} also indicate significant amount of $sp^2$-bonding. The origin of this shell is unclear: it may result from severe thermochemical treatment employed for separation of NDs from meteorite and thus represent damaged diamond surfaces and/or be a secondary material precipitated from the acids. Alternatively, they may reflect deviation of morphology of small grains of meteoritic NDs from assumed cuboctahedron shape. However, high resolution TEM studies complemented by DFT modeling of synthetic NDs grains showed well-developed faceting even for the grains less than 2~nm, although the $sp^{2+x}$ shell is always present \cite{stehlik2016high, shery2018size}. We note also that very small (1.6~nm in diameter) meteoritic NDs may possess intense photon emission of the Si-V defect \cite{vlasov2014molecular}; the fact indicating fairly perfect ordering of the nanograin lattice and little contribution of surface states. 

An important related issue is the presence and eventual
influence of surface functional groups on ND grains on the results of our
calculations. As noted above, the environment of formation of NDs in space
remains debatable, and chemical treatment complicates discussion of their
surface chemistry. In the case of NDs in outer space, the presence of
significant concentrations of surface-bound N and O is not very plausible.
However, hydrogen is almost certainly present, as suggested both by direct
astronomical observations of C-H vibrations at 3.43 and 3.53~$\mu$m assigned
to NDs \cite{Guillois_1999, van2002nanodiamonds} and unusual isotopic
composition of hydrogen in meteoritic NDs, possibly inherited from their
formation \cite{virag1989isotopic}. However, the extent of the surface
hydrogenation is yet uncertain; see discussion in Shiryaev, \emph{et al.}
\cite{shiryaev2015photoluminescence}. The influence of surficial hydrogen on
the results of our calculations requires additional investigation, but we
recall again that even for 1.6~nm ND grains influence of surface states appears
to be minor. 

The current work provides atomistic explanation of a remarkable phenomenon --
release pattern of heavy NGs implanted into ND grains is bimodal and is
unimodal for He. We show that the release patterns are satisfactorily explained
by such peculiarities of the behavior of the implanted ion in a nanograin, as
implantation depth, in-grain diffusion and thermal evolution of the grain
structure heated by the ion impact. Besides general interest for understanding
of size effects in interaction of ions with matter, this work is relevant to
carbon dust evolution in space. Two main scientific tasks are important: (i)
origin of isotopically different NG components and (ii) presence of different
ND populations, differing in origin, size and/or morphology. 

Xenon belonging to two of the three main NG components in NDs -- P3 and HL -- releases in two main peaks. It is usually assumed that the low temperature peak is dominated by the P3 gases and the high temperature one by the HL gases. The P3 gases appear to reside in traps with broad range of activation energies \cite{Huss-Met-1994-II} and are readily lost during thermal metamorphism of the parent bodies of meteorites. Investigation of size-separated NDs revealed that the ratio of P3/HL decreases in small ND grains, but the absolute concentrations of both components increase with the size \cite{Verchovsky-Science-1998, fisenko2004noble}.  

Although the present work does not reject the possibility of association of the
P3 component with species adsorbed in surficial $sp^2$-containing shells, it
shows that monoenergetic implantation alone nicely explains both components. In
this case, both low- and high-temperature peaks should be present in a given
irradiated population of NDs, implying that if implanted, the P3 and HL gases
should contribute to the whole release pattern. In fact, Huss and Lewis \cite{Huss-Met-1994-II} and Huss, \emph{et al.} \cite{huss2008noble} suggested that an important
fraction of HL gases -- ``labile HL'' -- is released roughly simultaneously
with the P3 component and thus was lost during thermal metamorphism of parent
meteorite and/or due to heating in the nebula. In contrast,
Koscheev, \emph{et al.} \cite{Koscheev-Nature-2001} and Fisenko, \emph{et al.} \cite{fisenkoa2020meteoritic} proposed that it
is the P3 component, which dominates both low- and high-temperature release
peaks. Taken together with the hypotheses discussed above, our
work indicates that calculation of the Xe-HL isotopic composition should
include corrections both for the high-temperature P3 release and for
preferential enrichment in heavy isotopes at these steps due to isotopic
fractionation. Derivation of the isotopic composition of the ``pure'' HL
component remains nevertheless difficult since it is likely a product of more
than one nucleosynthetic process.

The claim that the P3 component should have a high-temperature
component also requires the assumption that isotopes of other elements should
be implanted. In particular, the contribution of extinct iodine isotopes --
$^{129}$I and, perhaps, $^{128}$I, is invoked for the explanation of the Xe-P3
composition (see Gilmour, \emph{et al.} \cite{Gilmour-GCA-2005, gilmour2016xenon} for details).
Contribution of surface-bound $^{129}$I can be ruled out since the energy of
the $\beta$-decay is far too small to drive the Xe recoil into an ND grain.
Subsequently, implantation of radioactive iodine isotopes should have occurred.
In recent work, Gilmour, \emph{et al.} \cite{gilmour2016xenon} suggested that a fraction of Xe
released at high temperatures was also produced in situ from iodine. Although
we have not explicitly analysed the behaviour of implanted iodine, we believe
that the similarity in masses between I and Xe should lead to qualitatively
similar behaviour of these elements.

The decrease of the P3/HL ratio in smaller ND grains is; however, more difficult to explain unambiguously. In small grains almost all implants will be in close proximity to the surface and direct application of our modeling would suggest higher fraction of the low-temperature release. However, distortion of shape of the smallest grains \cite{shery2018size} influences curvature of the facets and thus may strongly influence diffusion paths of an implant. Variations of the curvature as well and interaction of the implant with $sp^{2+x}$ surface shell may also be responsible for broad range of bonding energies of the P3 traps. In the same time, there is an ample evidence that NDs from every meteorite represent a mixture of several independent populations of grains \cite{russell1996carbon, Verchovsky-Science-1998}, which may have very different sources. 

Examination of Figures~\ref{Figure9} and \ref{Figure11} might
give some hints for the origin of the high-temperature Xe-P6. The highest
release temperatures were observed for few events of very deep Xe implantation.
Examination of size-separated NDs suggests that the P6 component appears to be
present mostly in coarse grains \cite{Gilmour-GCA-2005}. Our calculations
addressed only relatively small ND grains with a diameter of 3.9~nm.
Examination of larger grains is computationally very expensive and was not
performed, but it is reasonable that for a given implantation energy and larger
grain size, a larger fraction of the ions will stop at greater depths. The
release temperature of these ions will also tend to cluster at higher values. 

Last but not least, a result of our calculations is a confirmation of the
stability of a Xe-V point defect in a ND grain. In annealed ion-implanted
macroscopic diamond this defect possess rather strong infra-red
photoluminescence (lines at 793--794 and 811--813~nm) when excited by visible
light \cite{deshko2010spectroscopy}(and references therein). In nanoparticles
the annealing required for conversion of the implant to the favourable
configuration occurs during the implantation process itself. Together with
luminescence of the silicon-vacancy (Si-V) defects
\cite{shiryaev2015photoluminescence}, the Xe-V complex opens a new possibility
for eventual observation of NDs in stellar environments.

\section{Conclusion} 

In this work we developed a molecular dynamics approach to address the
important unresolved question of the origin of the unimodal and bimodal thermal
release patterns of NGs from meteoritic NDs. Our technique employs a large
number (circa $10^4$) of small MD simulations to calculate the thermal release
pattern of implanted NGs in ND and provides detailed atomistic insight. We
reproduce the known experimental profiles for He and Xe, and propose that the
position and structure of the noble-gas defect are responsible for the unimodal
and bimodal patterns, respectively. In the case of Xe, we show that the low
temperature component is associated with shallow implantation on the \{100\}
faces, while the Xe-HL component is a mixture of deeply buried defects and
shallow implantation on the \{111\} faces. Although our
simulation methodology does not capture every single experimental detail, it
appears that a relatively straightforward description of the system captures
the key features. Areas of extension in future work include the addition of
hydrogen, modelling of the sp$^{2+x}$ shell, and improved potentials to
describe the energetics of defect migration and the \{111\} surface reconstruction. 

The MD methodology is natural fit for cloud-computing facilities and can be
easily extended to other systems where NGs are found in pre-solar grains. In
NDs, obvious directions for future work include unimodal release  in $^{3}$He,
$^{20}$Ne, $^{21}$Ne and $^{22}$Ne and bimodal behaviour of $^{38}$Ar,
$^{36}$Ar, $^{84}$Kr and $^{86}$Kr. Success in replication of the isotopic
effects may serve as a tool to investigate computationally implantation
processes of short-living isotopes; for example, spanning the whole range of
known Xe isotopes from $^{108}$Xe to $^{148}$Xe. 

Aside from NDs, there are several other materials relevant for interstellar
dust studies where our MD methodology can be applied. These include silicon
carbide (SiC) and graphite \cite{Ott-Nature-1993, Ott-SpaceScienceRew-2007,
Ott-Geochemistry-2014, Davis-ProcNatAcSci-2011, Amari-Nature-1990,
Amari-Nature-1993}. These pre-solar materials have high melting points and
therefore are suitable for our Arrhenius-based approach. Ballistic stage of ion
implantation in nano- and micron-sized C and SiC grains was studied using
Monte-Carlo-based approach (SRIM) \cite{verchovsky2003ion}. However, these
calculations cannot address grain heating, annihilation of defects and many
other effects. Therefore, whereas the range of ions is rather accurately
predicted, the overall impact of the implantation on nanograins remains
elusive. A number of successful atomistic simulation studies address
high-temperature effects in SiC \cite{Fujisawa-Carbon-2019}, graphite and
carbide-derived carbons \cite{deTomas-Carbon-2017, deTomas-APL-2018}. These
structures provide a natural starting point for further atomistic simulations
to reveal the astrophysical secrets of NGs in pre-solar grains.

The main findings of our work are summarized as following;
\begin{itemize}
	\item[(i)] Upon heating, monoenergetic Xe ions implanted into an ND grain will be released in two distinct peaks at markedly different temperatures. 
	\item[(ii)] Implanted monoenergetically He ions will be released as a single broad peak with an onset shifted to higher temperatures relative to Xe.
	\item[(iii)] Isotopic fractionation favouring heavy isotopes accompanies high-temperature steps.
	\item[(iv)] Differences in the behaviour of implanted NGs are explained by diverse types of formed lattice defects and diffusion paths. 
	\item[(vi)] Exact release pattern of the implanted NGs depends on the morphology of the grain, the position of the implant in the grain, and the peculiarities of the impact.
	\item[(vii)] Our results suggest that the implantation scenario can explain both the P3 and HL components of Xe in NDs. In addition, both components should possess low- and high-temperature release peaks. The absence of a low-temperature peak for the HL and, possibly, for the P6 component is due to post-implantation annealing of the grains.
\end{itemize}

\section*{Acknowledgements} 

We gratefully acknowledge many helpful discussions with Ulrich Ott.  NAM acknowledges fellowship FT120100924 from the Australian Research Council.  Computational resources were provided by the Pawsey Supercomputing Centre with funding from the Australian Government and the Government of Western Australia.


\pagebreak
\widetext

\onecolumngrid

\begin{center}
\normalsize{~\\ Supplementary Information} \\~\\~ \Large{Molecular dynamics approach for predicting release temperatures of noble gases in pre-solar nanodiamonds}\\[0.8cm]
\end{center}

\setcounter{section}{0}
\setcounter{equation}{0}
\setcounter{figure}{0}
\setcounter{table}{0}
\makeatletter
\renewcommand{\thesection}{S\arabic{section}}
\renewcommand{\theequation}{S\arabic{equation}}
\renewcommand{\thefigure}{S\arabic{figure}}

In this Supplementary Information, we provide further information on:

\begin{itemize}

    \item Helium--carbon and xenon--carbon interactions.
    \item Annealing of pristine ND
    \item Movies of helium and xenon release processes at high temperature. 

\end{itemize}

All the equations, sections and figure in Supplementary Information are labeled
with the prefix ``S'' to be distinguished from those that appear in the body of
the paper.\\[0.5cm]

\twocolumngrid

\section{He--C and Xe--C interactions}
\label{AppendixA}

\begin{figure}[!b]
\centering
\includegraphics[width=0.38\textwidth]{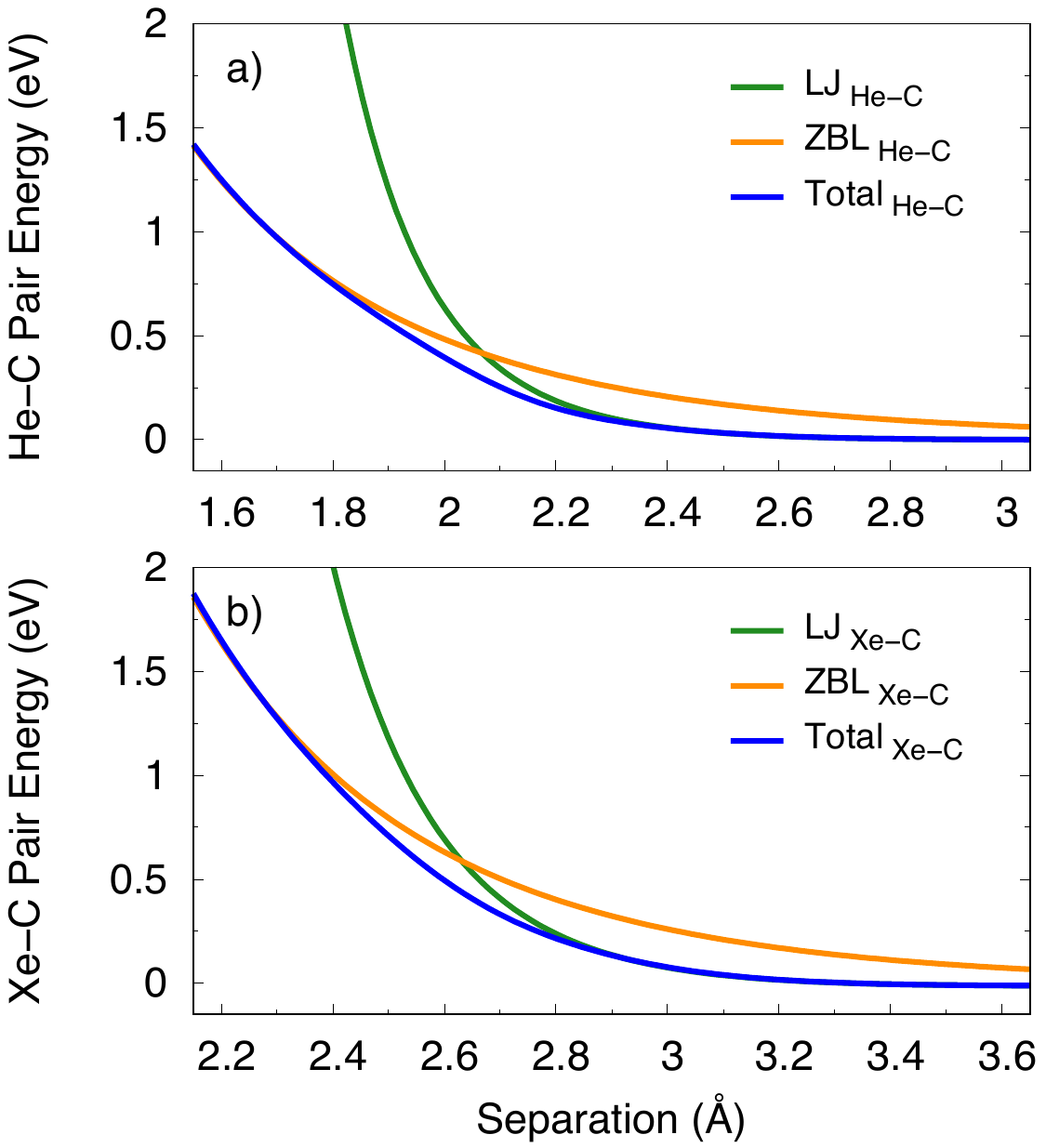}
\caption{ Pairwise interaction energies (blue line) for (a) He--C, and
(b) Xe--C. At close approach the interaction is pure ZBL (orange line),
while at distances around equilibrium and greater, a Lennard-Jones expression
is used (green line). }
\label{FigureS1}
\end{figure}

\begin{figure*}[!t]
\centering
\includegraphics[width=0.89\textwidth]{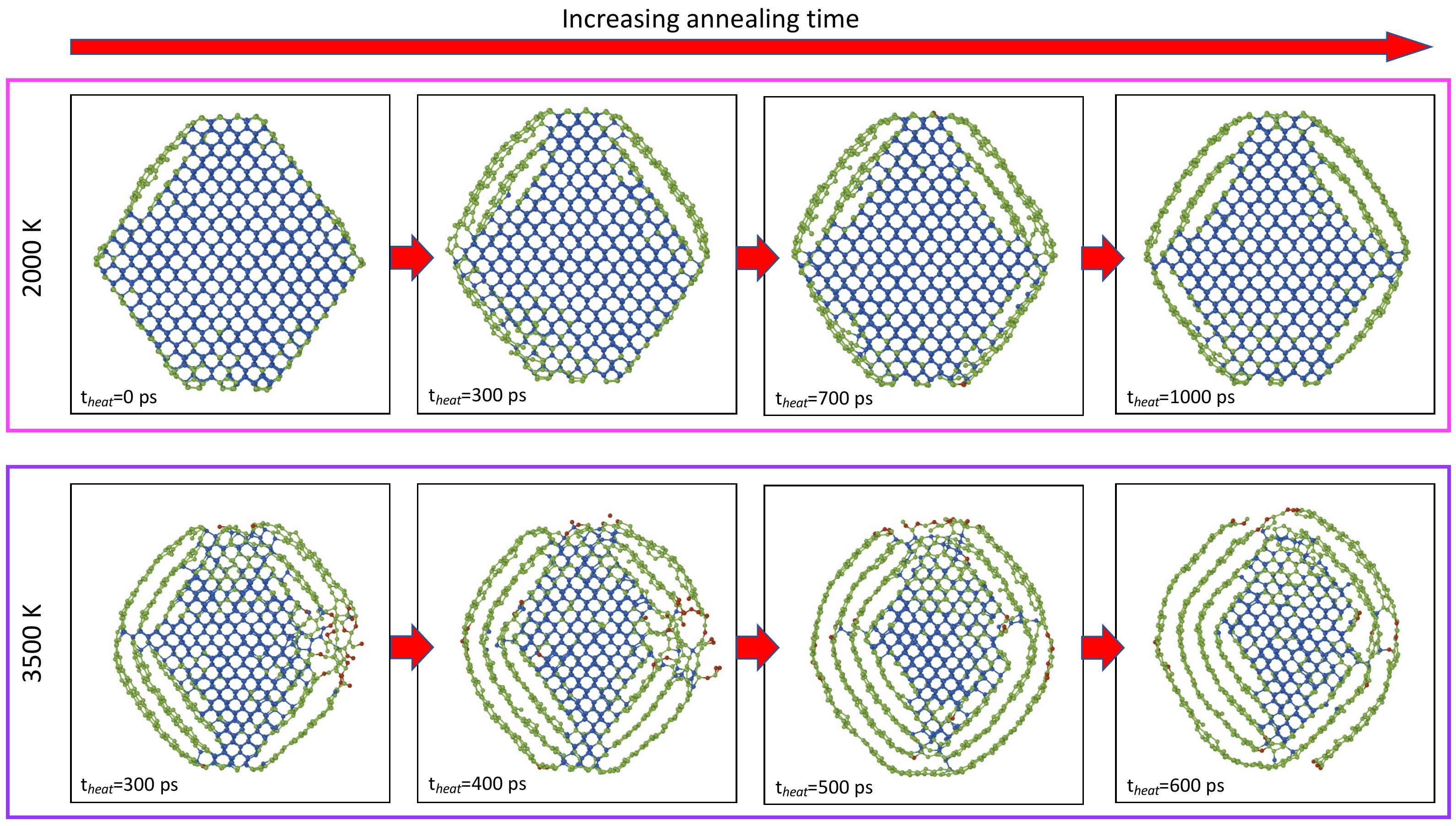}
\caption{Cross-sectional snapshots for the time evolution of 
pristine ND for annealing at 2000~K (upper panel) and 3500~K (lower panel)
showing how graphitization initiates first on the \{111\} faces, followed
later by the \{100\} faces. The color codings and slice details are the
same as Figure~\ref{Figure2}. }
\label{FigureS2}
\end{figure*}

As described in the Methodology section of the main text, for the
He--C and Xe--C interactions, we use the standard
Ziegler-Biersack-Littmark (ZBL) potential coupled with a Lennard-Jones (LJ)
potential. The LJ potential has the form 
\begin{equation} 
	U_\mathrm{LJ}(r) =
	4 \varepsilon \left\{ \left( \frac{\sigma}{r} \right)^{12} - \left(
	\frac{\sigma}{r} \right)^6 \right\} , 
\end{equation} 
where for He--C
interactions $\varepsilon_\mathrm{He-C}$=0.0013~eV and
$\sigma_\mathrm{He-C}$=2.98~\AA\ \cite{Nguyen-Langmuir-2008} and for
Xe--C interactions $\varepsilon_\mathrm{Xe-C}$=0.0114~eV and
$\sigma_\mathrm{Xe-C}$=3.332~\AA\ \cite{Simonyan-JCP-2001}. Following
\cite{Buchan-JAP-2015} and \cite{Christie-Carbon-2015}, the total
interaction energy is expressed as 
\begin{equation} U_\mathrm{Total}(r) = \Big[
	U_\mathrm{ZBL}(r) \times    f(r+\delta) \Big] + \Big[ U_\mathrm{LJ}(r)  \times
	(1-f(r-\delta))\Big] , 
\end{equation} 
where $f(r)$ is a Fermi-type switching
function, $\delta_\mathrm{He-C}$=0.09~\AA\ and $\delta_\mathrm{Xe-C}$=0.07~\AA\
\cite{Fogg-NIMB-2019}. The Fermi function is given by 
\begin{equation} f(r) =
	\Big[ 1 + \exp{ \big(b_F(r-r_F) \big) }\Big]^{-1} \label{Fermi} 
\end{equation}
where $b_F$ controls the sharpness of the transition and $r_F$ is the cutoff
distance; these parameters are chosen manually to ensure smoothness. For Xe--C
we use $b_F$=8~\AA\ and $r_F$=2.7~\AA, the same as our recent implantation
studies \cite{Shiryaev-SciRep-2018,Fogg-NIMB-2019}, while for He--C we use
$b_F$=9.2~\AA\ and $r_F$=2.2~\AA. Figure~\ref{FigureS1} illustrates the He--C
and Xe--C interaction energy covering the ZBL and LJ regimes. Further
discussion and full details of the interpolation process are provided in 
Buchan, \emph{et al.} \cite{Buchan-JAP-2015} and Christie, \emph{et al.} \cite{Christie-Carbon-2015}.

\section{Annealing of pristine ND}
\label{AppendixB}

As described in the main text, the the simulation temperatures are mapped onto their experimental equivalent using the Arrhenius relation. Calibration is performed using the graphitization of NDs into C onions as a reference point. Figure~\ref{Figure2} in the main text illustrates cross-sectional snapshots of the annealed ND for eight different temperatures. Panel (a) shows that at 1000~K there is insufficient temperature to graphitize the ND on this timescale, while at 1500 and 2000~K (panels (b) and (c)) some of the bonds on the \{111\} faces are broken and graphitic shells form. The reason this occurs is that $sp^3$--C hybridized atoms near the surface can easily rearrange into graphite with $sp^2$--C hybridization \cite{Kuznetsov-ChemPhysLett-1994, Wang-PRL-2000, Pantea-JAP-2002, Barnard-DRM-2003}. At higher annealing temperatures of 2500 and 3000~K (panels (d) and (e)), graphitization proceeds further inwards, creating structures with a diamond core surrounded by a graphitic shell. This core-shell structure is in good agreement with previous experimental and simulation studies \cite{Kuznetsov-ChemPhysLett-1994, Kuznetsov-Carbon-1994, Tomita-Carbon-2002, Los-PRB-2005, Qiao-ScriMat-2006, Brodka-DRM-2006, Brodka-JMS-2008, Ganesh-JAP-2011}. At 3500~K (panel (f)), a pure carbon onion (or concentric fullerene) is obtained and only a few small number of atoms ($\sim$1\%) have been lost via evaporation. At this point we have identified the temperature at which the experimental result is reproduced on the timescale that is affordable in the simulation. At the slightly higher temperature of 4000~K (panel (g)), around 8\% of the atoms evaporate and the C atom develops a hollow core; similar structures are seen in Figure~8(e) of Lau, \emph{et al.} \cite{Lau-PRB-2007}. At the highest temperature of 4500~K (panel (h)), the ND is completely destroyed and more than 29\% of atoms are evaporated, creating a large, disordered fullerene.

The time-evolution of the graphitization process is shown in Figure~\ref{FigureS2}
for two different annealing temperatures. The upper panel shows data at 2000~K,
where as time increases, the \{111\} faces graphitize first, following later by the 
\{100\} faces. At the higher temperature of 3500~K (lower panel), a similar sequence 
occurs, with the addition of an unravelling process which is facilitated by the 
development of spirals.

\section{Supplementary Movie}

Time evolution of He (first part) and Xe (second part) release processes from $\sim$3.9~nm ND during the high-temperature annealing (as discussed at the Figure~\ref{Figure4} in the main text). The thickness of the slice is 1~nm. The pink lines indicate the He and Xe trajectories, and the He and Xe are shown as an orange circle. C atoms are shown as red, green and blue circles with $sp$--C, $sp^2$--C and $sp^3$--C hybridizations, respectively. 

\twocolumngrid

\bibliography{references.bib}
\end{document}